\documentclass[
 superscriptaddress,
 amsmath,amssymb,amsfonts,
 aps,
 prb,
 twocolumn, 10pt,
 longbibliography,
 floatfix
]{revtex4-1}
\usepackage[utf8]{inputenc}
\usepackage{graphicx}
\usepackage[sf]{subfigure}
\usepackage{color}
\usepackage{fouriernc}
\usepackage[dvipsnames,svgnames,x11names,hyperref]{xcolor}
\usepackage[low-sup]{subdepth}
\usepackage{MnSymbol}
\usepackage{multirow}
\usepackage{outlines}
\usepackage{etoolbox}
\usepackage{enumitem}
\bibpunct{[}{]}{,}{n}{}{}

\definecolor{linkcolor}{RGB}{6,69,173}
\usepackage[T1]{fontenc}
\usepackage[utf8]{inputenc}
\usepackage[colorlinks=true,
            linkcolor=linkcolor,
            urlcolor=linkcolor,
            citecolor=linkcolor,
            unicode,
            pdfencoding=auto]{hyperref}

\usepackage[
]{physpack}

\begin{document}

\title{Fast-Scrambling and Operator Confinement Using an Auxiliary Qubit}
\author{Joseph C. Szabo}
\affiliation{Department of Physics, The Ohio State University, Columbus, Ohio 43210, USA}

\author{Nandini Trivedi}
\affiliation{Department of Physics, The Ohio State University, Columbus, Ohio 43210, USA}

\begin{abstract}
We introduce a minimal model for realizing a fast-to-slow scrambling transition mediated by an auxiliary central qubit (c-qubit). The c-qubit is coupled to a spin-$1/2$ Ising model with local Ising interactions and tunable c-qubit-spin coupling. Each spin becomes next-nearest neighbor to all others through the c-qubit, 
which mediates effective all-to-all interactions. 
As the interaction with the c-spin increases, we find a surprising transition from super-ballistic scrambling and information growth to continuously restricted sub-ballistic entanglement and operator growth. This slow growth occurs on intermediate timescales that extend exponentially with increasing coupling and system size, indicative of logarithmic entanglement growth. We find that in the slow-scrambling regime, the c-qubit Ising interaction allows commuting operators to grow support on all sites rapidly, while operators orthogonal to the interaction become echoed out. This projects local operators to lie in a restricted subspace and prevents extensive operator entanglement growth. We provide exact dynamics of small systems working with non-equilibrium, effective infinite temperature states, and additionally contribute analytic early-time expansions that support the observed rapid scrambling to quantum Zeno-like crossover. Tracing out the central qubit provides a unique translation from the full, closed unitary dynamics to a simple open system construction consisting of a typical spin-chain with hidden qubit degree of freedom.
\end{abstract}
\date{\today}

\maketitle

\section{Introduction} 
Operator scrambling and entanglement entropy spreading are unambiguous discriminators of purely quantum mechanical nonequilibrium dynamics; fascinating properties underlying quantum thermalization, dynamical phase transitions, and topological order~\cite{tee_kitaev, tee_wen, Hastings_2007, entanglement_spectra_Haldane, entSpec_topoPhase_pollman, Heyl2018, Dag2020}. In the Heisenberg picture, quantum operator scrambling details how initially localized operators propagate over spatiotemporal degrees of freedom due to noncommutative many-body interactions. In the complementary Schrodinger picture, entanglement entropy captures growing information complexity: from initially classical states to those with nontrivial entangled structure. Quantum information dynamics bridge both theoretical and experimental communities as primary measures for quantum complexity and expressivity ~\cite{landsman2019, joshi2020, blok2021, google2021, china_scramble_2022}. These concepts combined with quantum simulation/circuit devices have coalesced into many enriching, recent experiments ~\cite{mi2021information, blok2021, mi2022time, zhu2022observation}. The accelerating pace of results and drive to continually advance the corresponding theory extend these successes to further research at the intersection of quantum chaos, thermalization, and computability, extending from qubits to black-holes and quantum gravity~\cite{entropy_srednicki, Sekino_2008, Lashkari2013, Maldacena_2016, Maldacena2016, Chen2017, Bohrdt_2017, Rigol2008, Chen2017, lewis-swan2019}.



The primary research thrusts among the quantum information dynamics community fall along the lines of uncovering the minimal mechanisms behind myriad information dynamical phases and understanding the fate of quantum to classical thermalization.
In studying generalized quantum information dynamics, there are typically two disparate perspectives: closed and open quantum systems. Closed quantum systems exhibit rich scrambling physics ranging from frozen~\cite{Swingle2017} to fast~\cite{Lashkari2013} dynamics, with the typical questions relating to how well-preserved is such physics under driving and dissipation contributions from an external environment~\cite{zeng2017prethermal, moessner2017equilibration, Touil2021, zanardi2021information}. The environment is oftentimes reduced to a memoryless, effective Markovian description, which hinges on assumptions including weak-coupling and a separation in the timescales associated with system and environment~\cite{mazzola2010phenomenological, cohen1998atom}. Though solving for the exact dynamics for a full complex environment is beyond the capabilities of current devices, taking into account the structure and interaction with the environment poses interesting research questions: what is the fate of entangled information within the system, how can a structured environments drive effective interactions and information dynamics, and how does the environment serve as a probe in an information theoretic/entropic capacity?

A simple avenue for exploring the impact of a structured quantum environment is by considering composite, unitary models. Focusing on a particular \textit{subsystem} of a full closed quantum system and tracing over the additional degrees of freedom (DoFs), then termed the \textit{environment}, captures the subsystem's effective dynamics/interactions. This is a popular focus of study as it provides an open quantum system perspective for the subsystem and allows full consideration of the environment's structure and interaction topology. This construction allows us to specifically evaluate how variable structured environments impact the overall quantum information dynamical phase as expressed by the underlying subsystem. Previous work considered the validity of Markovian assumptions provided variable system-environment coupling, and here we are looking to add an information scrambling perspective. Considering a system-environment construction in this manner directly applies to those codes/models investigating the information physics of auxiliary bits or those systems with inherent auxiliary DoFs such as mechanical or optical modes~\cite{chu2017quantum, mirhosseini2020superconducting, kok2007linear, stockill2017phase}.

Studying composite systems in this manner provides an interesting framework extending current quantum information dynamic research. Significant recent results focus on the range of interactions, the speed and nature of information propagation, the role of inherent symmetries, and the effect of emergent symmetries in the cases of many-body localization, Floquet periodic driving, etc. The same phenomena can be similarly cast as an environment mediated effect. The tunability of the system-environment network topology and the inherent environment structure and interaction symmetries then allows for systematic investigation into the particular contribution on the overall dynamics.


In this paper we explore these aforementioned questions by considering the simplest environment extension; an auxiliary central qubit (c-qubit) coupled to 1-d chain system of interacting spin-$1/2$ (qubit) objects. Tracing out the c-qubit and considering the dynamics of the 1-d system provides a translation from a full, unitary model, to an effective long-range, non-Hermitian spin chain with a hidden qubit degree of freedom. This provides a single long-range quantum channel for transmitting information but at the same time imposes a shared two-fold DoF across all spins. Though only a small addition to well-understood nearest-neighbor qubit model, we observe an abundance of exciting repercussions. 

The system-environment coupling expresses various regimes: in the weakly coupled regime, the c-qubit provides little feedback and acts as a free channel for information to pass unimpeded; while when strongly coupled to the low dimensional qubit environment, the c-qubit acts as a strong drive and imposes an effective hidden symmetry on the underlying spin system and generates \textit{disorder-free localization}. We liken the physics observed here to that seen in systems undergoing quantum measurement or strong Floquet driving. Considering the c-qubit as a hidden degree of freedom provides unique insight into how the quantum scrambling dynamics of the underlying spin chain maps to an extended unitary model provided one additional qubit. Central qubit or a higher dimensional qudit/cavity/register are popular theoretical and experimental tools for providing non-invasive many-body measurements~\cite{otocs_xxladder_dag, swingle2016, DelRe2022, singal2022implementation}, evaluating Hermitian and non-Hermitian response~\cite{geier2022, DelRe2022}, generating effective interactions~\cite{molignini2022}, and studying the fundamentals of decoherence and information transport~\cite{stricker2020experimental, wong2022quantum}.

\begin{figure*}[htb!]
    \centering
    \includegraphics[width=0.995\textwidth]{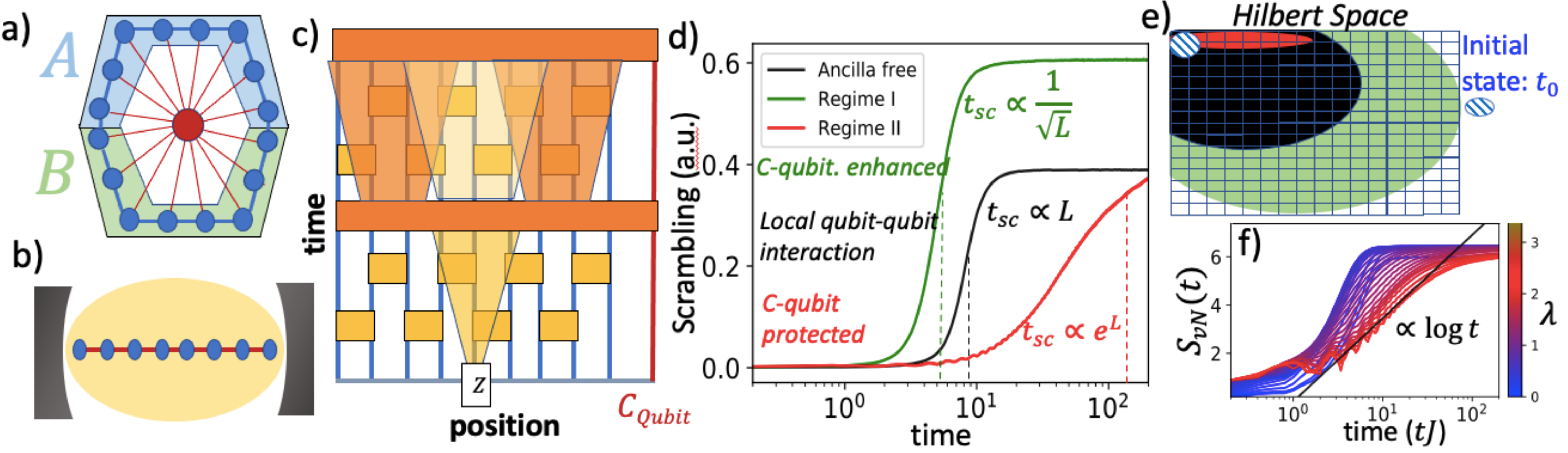}
    \caption{(a) Illustration of the ring-star model achieved in (b) optically dressed trapped-ion experiments and (c) through nonlocal unitary gates in a circuit realization. (d) Central qubit mediated dynamics: local interaction (black), c-qubit induced fast scrambling (green), and c-qubit inhibited scrambling (red). Information scrambling approximated as wavefunction spread in Hilbert space or $\sim e^{S_{vN}(t)}$ to show dramatic prethermal-like approach to full thermalization. (e) Qualitative depiction of (d) or how a well defined initial state grows to fill state/operator Hilbert space with a varying rate depending on dynamical phase. (f) von Neumann entanglement entropy $S_{vN}(t)$ following a quench from polarized state $|+y\rangle$ for the ring-star Ising model under varying $\lambda$.}
    \label{fig:cartoon}
\end{figure*}

Here we particularly focus on the dual effect of a tunable central qubit by investigating the operator and entanglement growth in a nonintegrable, ring-star Ising model. The model includes homogeneous spin-spin interactions in a mixed magnetic field. We find an extremely surprising fast-to-slow quantum information spreading transition that occurs due to the nonlocal and coherent nature of the c-qubit. We summarize this result in Fig.\ref{fig:cartoon}(d), where in the weakly coupled regime (regime I), the central qubit mediates rapid scrambling with a timescale that decreases with system size (green, upper curve in). In the strong coupling regime (regime II), the scrambling time increases exponentially with system size (red, lower curve). The mechanism behind this transition is the interplay between the noncommuting, extensive c-qubit Ising interactions and transverse field $h_c$. The metric here presented for scrambling is $e^{S_{vN}(t)}  / 2^{L+1}$, which provides a measure of the span of the quantum wavefunction throughout the full Hilbert space. As we detail in what follows, in the strong coupling regime the central qubit rapidly saturates its entanglement with the surrounding spin-chain environment and becomes strongly driven by this extensive interaction. This strong interaction rapidly aliases operators orthogonal to the central qubit Ising interaction on the central qubit and even more surprisingly within the spin chain. The long lifetime of states and operators that commute with the central Ising interaction leads to slow multi-particle entanglement growth and operator complexity.

Our work agrees with previous research that finds an extensively scaling, nonlocal interaction leads to rapid scrambling, where the rate increases with system size~\cite{Bentsen2019, Li2020, Belyansky2020}. At the same time we find a surprising limit where the purely quantum nature of the c-qubit imparts a coherent effect that slows operator decoherence/entanglement and subsequent spreading. This phenomena mirrors what is seen in strongly driven Floquet systems, where periodic driving can impart an effective symmetry in all eigenstates and leads to prethermalization and correspondingly slow entanglement spreading. We liken the projective action of the central qubit in this time independent Hamiltonian to the quantum Zeno effect where quantum measurement leads to a ballistic to sub-ballistic entanglement growth transition. Here the c-qubit imparts a highly nonlocal effect on operator projection in contrast to a local purification/disorder network that redefines local spreading dynamics. We illuminate this c-qubit physics by examining the growth of out-of-time-order correlators (OTOCs) and the von Neumann entanglement entropy for sufficiently high-energy initial product states.

\section{Scrambling Metrics}
Many recent works have made significant progress on establishing the family of scrambling dynamics that occur in various lattice models and geometric random circuit designs, as characterized by the growth of OTOCs. The OTOC generically given as 
\begin{equation}
    C_{VW} = \langle [\hat{W}(j,t), \hat{V}(i,0)]^\dagger [\hat{W}(j,t), \hat{V}(i,0)] \rangle,
    \label{eq:OTOC_full}
\end{equation}
examines how an initially prepared unitary operator $\hat{V}$ on site$-i$ commutes after Heisenberg evolution with operator $\hat{W}$ after time $t$ (here assumed a local operator on site$-j$). The operator spreading picture is unique to quantum systems, where in working with pure states, no information is truly lost but transforms into many-body degrees of freedom that become increasingly inaccessible provided control over an initial localized region.

Studying OTOCs and the timescales associated with scrambling dynamics provides a conjugate perspective as compared to entanglement entropy measures and transitions. Where OTOCs and specifically infinite temperature OTOCs examine the light cone established by Heisenberg evolution and depend more strongly on the commutivity graph, entanglement entropy examines how the wavefunction over a bipartition of Hilbert space spreads throughout. Here we specifically focus on the von Neumann bipartite entanglement entropy given as 
\begin{equation}
    S_{vN} = - \sum_k \lambda_k \log(\lambda_k),
\end{equation}
where $\lambda_k$ are eigenvalues of the reduced density matrix $\rho_{A|B}$ (RDM) obtained by integrating out subsystem $\mathcal{A} \text{ or } \mathcal{B}$ with corresponding Hilbert spaces $\mathcal{H}_A, \mathcal{H}_B$. OTOCs and entanglement identify similar physics and previous colloquial conceptions of the two established quantum scrambling as a unifying framework behind them; where, scrambling represents the time for an OTOC between arbitrary sites to become $O(1)$ and entanglement entropy to become $O(L)$. In the case of OTOCs, this limit is not rigorous enough and only provides a best-case scenario for operators traversing the system rather than providing a timescale for nontrivial operator strings to span the system~\cite{Harrow2021} (extensive operator entanglement). Rigorous relationships between OTOCs and Renyi$-2$ entropy have been established~\cite{Zanardi2001, Yan2020, Styliaris2021} and special cases have been studied in particular optical Hamiltonians~\cite{Garttner2017, li2017, lewis-swan2019}.

Unitary scrambling physics generally falls into two categories: systems that thermalize rapidly and those that fail to do so. The former are known as fast-scramblers; ergodic systems typically with variable all-to-all range interactions that spread quantum information throughout the full Hilbert space in $t_{sc} \sim \log(N)$. Models such as Sachdev-Ye-Kitaev (SYK) and non-integrable infinite range Ising and XY models are known to exhibit fast-scrambling physics~\cite{KitaevKITP, Iyoda, Li2020, Belyansky2020, Bentsen_2019}. Systems that fail to thermalize with $t_{sc} \sim e^L$ are slow-scramblers, non-ETH obeying systems and candidates for highly coherent quantum information storage. There are multiple vectors through which non-ETH physics occurs: integrability, disorder-free localization~\cite{Hart2021, Halimeh2022}, quantum scarring ~\cite{turner2018, choi2019, serbyn2021, chandran2022}, and/or higher order exact or proximate conservation laws~\cite{prem2017, pai2020, feng2022}. The origins of much of this work stems from the dramatic quantum correlations observed in quantum simulation experiments.

Typical models accessible to simulation and experiment are semiclassical in nature with infinite or long-range interactions. These models exhibit the characteristic unitary scrambling features we detailed previously, yet continue to enrich the discussion with new puzzling results. In Lipkin-Meshkov-Glick (LMG) model or the Dicke model exhibit strict conservation of the total spin moment $\hat{S}^2$ such that the effective number of degrees of freedom is $\mathcal{O}(L)$, compared to $\mathcal{O}(2^L)$~\cite{lewis-swan2019, Alavirad2019, lerose2020}. These systems have been observed to spread information rapidly, while the complexity saturation value remains low. This is in stark contrast to fast scrambling models like SYK where infinite-range connectivity allows for rapid and complex quantum information scrambling. One immediately puzzling question is: how do long-range interactions, tending toward generating semiclassical behavior, compete with local chaotic quantum dynamics to allow a fast-to-slow scrambling transition? A complete understanding of quantum information physics not only hinges on understanding the unitary dynamic contribution to experimental results, but similarly understanding non-Hermitian processes. These are inherent to quantum simulation platforms and represent the an exotic next frontier for theory and experiment as we move toward fully expressive quantum circuits and computation.

More generalized quantum dynamical behavior has been explored in recent studies consisting of non-Hermitian operations: composite system-environment undergoing quantum measurement~\cite{Li2019, Skinner2019, Jian2020, Bao2020, Lavasani2021, Minato2022}, light-matter interactions~\cite{lewis-swan2019}, dissipative and driven systems~\cite{luschen2017, choi2020quantum, lenarvcivc2020, wybo2020}. The most extraordinary findings reveal that these non-unitary dynamics generate effective inter-system interactions and impose effective static long-lived symmetries: Floquet periodically driven systems are akin to various unitary scrambling phases.

\section{Model}
We consider the Hamiltonian for the c-qubit or ring-star Ising model: 
\begin{equation}
    H = \sum_{i=0}^{L-1} \lambda \sigma^z_{i}\sigma^z_{c} - J\sigma^z_i\sigma^z_{i+1} + h \sigma^x_i + g \sigma^z_i + h_c\sigma^x_c + g_c\sigma^z_c,
    \label{eq:star_local_Ham}
\end{equation}
where $\lambda$ represents the uniform spin-c-qubit interaction, and $J, h, g$ represent the much-studied nonintegrable mixed-field Ising model. Here we take $J = 1.0, h=h_c=1.05, g=g_c=0.45$ unless otherwise noted. This is a well characterized nonintegrable point for polarized state evolution and operator dynamics allowing us to benchmark the impact of the c-qubit~\cite{Belyansky2020, kim2013}. 

Previous works examined operator dynamics and entanglement growth in random unitary circuits (RUCs) applied in a star-graph network. RUCs on the star-graph generated OTOC dynamics that saturated at $t \propto \log(L)$~\cite{lucas2019quantum, Harrow2021}. In contrast, an interesting section of \cite{lucas2019quantum}, examines dynamics of time-independent, star-Ising Hamiltonians which consist of Ising interactions and a non-commuting field $h_c$ strictly on the central qubit. In contrast to fast scrambling RUC networks, the time-independent c-qubit Hamiltonian dynamics generate novel operator confinement on the c-qubit with a coherent lifetime $\tau \propto \frac{h^2}{\lambda L}$. This defines the time for significant operator weight to decohere into operator subspaces that no longer commute with the Ising interaction, subsequently allowing slow scrambling to all other sites. Provided this contrast between the fast-scrambling RUCs and the slow Hamiltonian dynamics, we ask: can the c-qubit coupling mediate rapid scrambling in a truly time-independent nonintegrable model and, if so, when/how does this picture break down?

\section{Analytic Insights}

Before discussing the exact numerics on the full nonintegrable ring-star Ising model, we first motivate the reasons underlying a dynamical transition from fast-to-slow scrambling and the corresponding mechanism. We first summarize typical operator growth in local spin models and then review the Ising star graph dynamics.

\subsection{Local Ising Model}
In the completely local Ising chain limit $\lambda = 0$, we can gain a heuristic understanding of the characteristic light cone spreading by performing an early-time expansion of the dynamical correlator $\langle \sigma^z_i(t)\sigma^z_j \rangle$. Using the early-time expansion of the Heisenberg evolution we can write $e^{i\hat{H}t}\sigma^z_ie^{-i\hat{H}t}$ using the Baker–Campbell–Hausdorff (BCH) expansion
\begin{multline}
    \\
    e^{-i\hat{H}t}\sigma^z_ie^{-i\hat{H}t} = \sum_m^\infty \frac{(it)^m}{m!}[H, S^z_i]_m \\
    \\
    [A, B]_m = [A, [A,B]_{m-1}] ; [A, B]_0 = B \\
    \\
    [H, S^z_i]_1 = it[hS^x_i, S^z_i] = -ithS^y_i \\
    \\
    [H, S^z_i]_2 = \frac{(it^2)}{2}[H, S^y_i] \\
    = \frac{-t^2}{2}(-\lambda S^z_0S^x_i + hS^z_i - gS^x_i - J(S^z_{i-1}S^x_i + S^z_{i+1}S^x_i)).
\end{multline}
Before continuing this expansion to higher orders we observe a trend in operator growth and can approximate the weight as
\begin{equation}
    [S^z_i(t), S^z_j] \propto \frac{t^{|i-j|}}{|i-j|!}\mathcal{O}(1)
\end{equation}
Naively, we then expect the OTOC, or squared commutator, to generically follow 
\begin{equation}
    C_{zz}(i-j,t) \sim \frac{t^{2|i-j|}}{(|i-j|!)^2}\mathcal{O}(1)
    \label{eq:local_otoc}
\end{equation}
This approximation is valid for times $t \sim |i-j|/v_B$, where $v_B$ is the characteristic butterfly velocity. Before this time, operator growth is suppressed with an exponent that grows with distance, and allows for optimized simulations using matrix product operator dynamics (MPO) by just keeping track of the operator wavefront~\cite{Xu2020}. Though many extensive theoretical works provide rigorous estimates on the form of operator growth for translationally invariant models, here we simply examine the asymptotic form. This exponentially growing operator weight with exponent like $r = |i-j|$ leads to the development of a linear light cone with butterfly velocity exactly calculated as $2eJ$ ($e$ being Euler's number) for $h/J > 1$~\cite{Chen_2021}. Following this light cone in the nonintegrable Ising model, any localized operator spans the full operator Hilbert space $\{\hat{X}, \hat{Y}, \hat{Z}, \hat{I}\}$ within region $r$. 

An interesting extension of the OTOC is the integrated OTOC (iOTOC), which is the integral over $r$ and the bipartite OTOC~\cite{Keselman2021, Styliaris2021}. Both provide a novel characterization of the operator complexity within a region $r$. This also provides an insightful way to then understand the corresponding entanglement dynamics for pure state wavefunctions with energy density $k_BT$ as in accordance with ETH the states and operators should exhibit ergodic equilibration. In the nonintegrable Ising model, an operator saturates the reduced Hilbert space $4^{|i-j|}$ after a time $t_{sat} = \frac{|i-j|}{v_B}$, so simply integrating over Eq.\ref{eq:local_otoc} provides an exponentially growing iOTOC for all $C_{vw}$. This operator growth then guarantees ballistically growing entanglement entropy $S_{vN} \sim \sum_{v,w}\log[\text{iOTOC}(v,w)]$. 

\subsection{Star-Ising model}
In the star limit, we set $J = 0$ and tune external fields $h, g$ to allow more expansive operator evolution. First working with only $\lambda \neq 0$, we then add complexity to arrive at the full, nonintegrable ring-star Ising model. For $|\lambda| > 0$, operator growth between leaves of the graph is trivial. The dynamics are exactly solvable as $[\sigma^z_i, \hat{H}] = 0$ for all sites $i$. Similarly, the commutivity graph representing the Hamiltonian and how local operators propagate is completely disconnected with vertices $\sigma^z_i, \sigma^z_L$ for all $i\in L$ and no bonds in between. Operators evolve simply under the central Ising interaction and the two-time correlator behaves as:
\begin{equation}
    \langle S^x_i(t) | S^x_i \rangle = \cos{2\lambda t}.
\end{equation}
And, again, exactly in this case, the OTOC goes as:
\begin{equation}
    C_{xx}(i, i, t) = 2\sin^2(2\lambda t),
\end{equation}
Though this limit admits a trivial result, it allows us to understand the key coherent property of the c-qubit qubit. We see that for all times, operators orthogonal to the Ising interaction $\lambda$, initially prepared on the leaves, propagate to the c-qubit after a time $t = \pi / \lambda$. Because all terms in the Hamiltonian commute with this interaction $[\sigma^z_i\sigma^z_0, \sigma^z_j\sigma^z_0] = 0$, the action of operator development from leaves to c-qubit does not decohere into nontrivial, orthogonal operators $\{\hat{X}, \hat{Y}\}$. Due to this coherence, operators oscillate between the initial node and the c-qubit. If the operator was initially prepared on the c-qubit, it fluctuates onto all nodes, but instead of becoming a many-body operator, it is a collective superposition of $L$ unique two-body operators $\sum_j \sigma^{x,y}_0\sigma^z_j \in \mathcal{H} (4^{\otimes L})$. This superposition, though nonlocal on a timescale $\tau = \mathcal{O}(1)$, has minimal operator entanglement and the complexity of operators strings is fixed to be a maximum of $2$. In this scenario, similarly, minimal entanglement entropy develops as the number of unique operator states is of $\mathcal{L}$ in a Hilbert space of $4^L$. Quench experiments on such systems with nonzero homogeneous magnetic fields will only permit half-system entanglement to grow like $\log{L}$ since we have effectively a large semiclassical spin system coupled to a single two-level qubit~\cite{szabo2022}.

The simplest extension we can make is to now introduce a transverse field on the c-qubit spin, $h_c \neq 0$. This was similarly analyzed in a disordered star-Ising system for small values of $h / \lambda L$~\cite{lucas2019quantum}. For $|h_c| > 0$, the model retains the same integrability, as any $\{z_1, ... z_L\}$ is an eigenstate of the system and will not evolve under unitary dynamics. We can then reduce the problem to solving for the evolution of the central qubit in a mixed $x-z$ field where the effective longitudinal strength given by $\sum_i \sigma^z_i$. Though fully solvable, we find illuminating operator dynamics in the infinite temperature limit. Solving exactly for $\langle \sigma^x_i(t) \sigma^x_i \rangle$ as was first provided in ~\cite{lucas2019quantum}, we observe how the coherent oscillation $\langle \sigma^x_i(t) \sigma^x_i \rangle = \cos(2Jt)$ for $h_c = 0$, evolves for increasing $h_c$ and leads to operator decoherence. Operator decoherence, in this sense, is that as auxiliary operators such as $\sigma^z_c \rightarrow \sigma^x_c, \sigma^y_c$, they then no longer commute with the c-qubit Ising interaction and operator weight then grows on sites $j \neq i$. The superposition of these growing dynamical correlations on sites $j \neq i$ similarly decreases the probability of finding $\sigma^x_i$ on site$-i$ after time $t$, leading to a decay in the autocorrelation function on site $i$.

\begin{figure}
    \centering
    \includegraphics[width=0.45\textwidth]{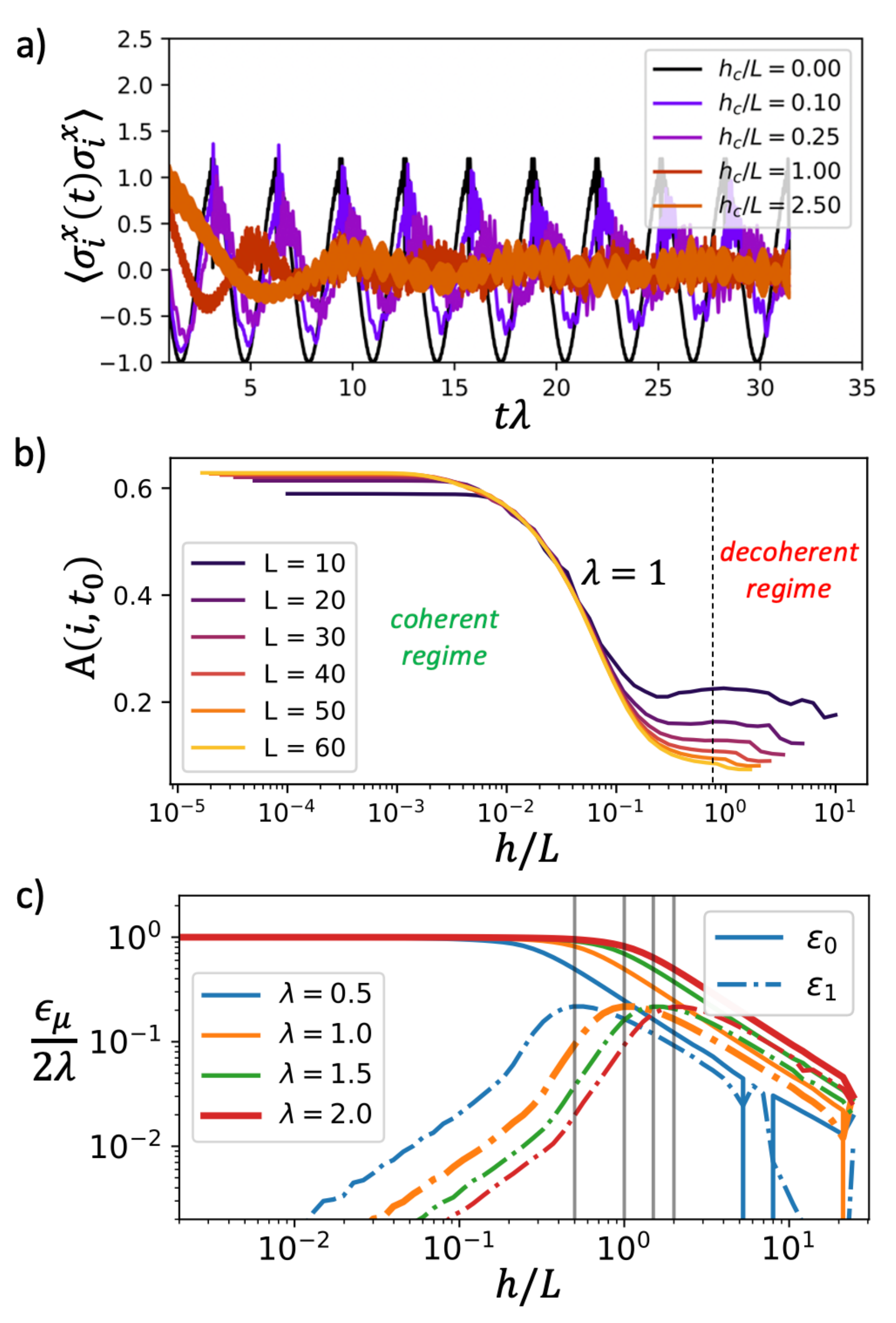}
    \caption{\textit{Autocorrelation $\langle \sigma^x_i(t) \sigma^x_i \rangle$: } (a) Eq.\ref{eq:star_exact} vs. time ($t\lambda)$ as a function of $h, L = 40, \lambda = 1.0$. 
    (b) Long-time average of (a) $A(i,t_0 = 200)$ as a function of $h$ and system size $L$. 
    (c) Curve fit of smoothed results in (a) consisting of coherent oscillating component with frequency $\epsilon_0$ and exponential decay parameter $\epsilon_1$. $\epsilon_0$ shown in the top, solid color palette remains equal to $2\lambda$ roughly until reaching the critical point $\lambda = h/L$. Vertical lines in (c) correspond to critical point $h_c = L \lambda = [0.5, 1.0, 1.5, 2.0]$.}
    \label{fig:star_analytic}
\end{figure}

For $\lambda L >> h$, it was shown numerically that early time operator dynamics go as 
\begin{equation}
    \langle \sigma^x_i(t) \sigma^x_i \rangle \sim \cos(2\lambda t)e^{-\sqrt{\frac{\pi}{2L}}\frac{h^2}{\lambda}t}
\end{equation}
using the memory matrix formalism~\cite{lucas2019quantum}. Regardless of whether $\lambda$ is a Gaussian random variable or homogeneous, we arrive at the same exponential dependence on system size. In the case of Gaussian random variables, this approximation was observed to be faster than the true decay rate. The key physics being that the $L-$site interactions with the c-qubit lead to an extensive coherence time, which we can evaluate explicitly by tracing over the set of eigenstates $\mathbb{Z}$:

\begin{gather}
    \nonumber \\
    \langle \sigma^x_i(t) \sigma^x_i \rangle = \frac{1}{2^L}\text{Tr}_{\mathbb{Z}}[\langle z_c, ... z_L | e^{i\hat{H}t}\sigma^x_i e^{-i\hat{H}t} \sigma^x_i | z_c, ... z_L \rangle] \\
    \nonumber \\
    = \langle z_c| \otimes \langle \mathbb{Z}_m | e^{i\hat{H}t}\sigma^x_i e^{-i\hat{H}t} | \mathbb{Z}_m \rangle \otimes | z_c \rangle \\
    \nonumber \\
    \hat{H} | \mathbb{Z}_m \rangle \otimes | z_c \rangle = e^{i(h_c\sigma^x_c + \lambda \sum_{i>0} z_i \sigma^z_c)t} |z_c \rangle \\
    \nonumber \\
    = \mathcal{I}\cos(2 \omega_{Z_m} t) + \frac{(h_c\sigma^x_c + \lambda Z_m \sigma^z_c)}{\omega_{Z_m}} \sin(2 \omega_{Z_m} t) |z_0\rangle.
\end{gather}
We use same simplified notation as \cite{lucas2019quantum}, where $\omega_\mathbb{Z_m} = \sqrt{h^2 + \lambda^2 Z_m^2}$, with $Z_m = \sum_{i>0} z_i$ and $\mathbb{Z_m} = \mathbb{Z}[m]$ is just $z-$eigenstate with magnetization $m \in [-L, L]$. As the set of states $\{z_1 ... z_L\}$ commute with the Hamiltonian, we reduce the Heisenberg evolution to that of a 2-level system and averaging over the respective density of states, which is simply a binomial distribution in the infinite temperature limit. Going back to the full evolution we then have
\begin{widetext}
\begin{gather}
    \nonumber \\
    \nonumber \langle \sigma^x_i(t) \sigma^x_i \rangle = \frac{1}{2^L} \sum_{z_0} \sum_m \left( \begin{array}{c} L \\ |m| \end{array} \right) \langle z_c | [\mathcal{I}\cos(2 \omega_{Z_m} t) + \frac{(h_c\sigma^x_c + \lambda Z_+ \sigma^z_c)}{\omega_{Z_m}} \sin(2 \omega_{Z_m} t)] \times \\
    [\mathcal{I}\cos(2 \omega_{Z^-_{m}} t) + \frac{(h_c\sigma^x_c + \lambda Z_- \sigma^z_c)}{\omega_{Z^-_{m}}} \sin(2 \omega_{Z^-_{m}}] t)]|z_c\rangle \\
    \nonumber \\
    \label{eq:star_exact} = \frac{1}{2^{L-1}} \sum_m \left( \begin{array}{c} L \\ |m| \end{array} \right) \cos(2 \omega_{Z_m} t)\cos(2 \omega_{Z^-_{m}} t) + \frac{(h_c^2 + \lambda^2 Z_m Z^-_m)}{\omega_{Z_m}\omega_{Z^-_m}}\sin(2 \omega_{Z_m} t)\sin(2 \omega_{Z^-_{m}} t).
\end{gather}
\end{widetext}
$\sigma^x_i$ flips the single spin state $z_i$ and leads to two unique frequencies $w_{Z_m}$ and $w_{Z^-_m}$, separated like $\lambda$ for $h = 0$, $Z^-_m = \sum_{j \neq i > 0} z_j - z_i$. $h_c = 0$ provides the exact, non-scrambling result $\langle \sigma^x_i(t) \sigma^x_i \rangle = \cos(2\lambda t)$, and for $\lambda = 0$ we have $\langle \sigma^x_i(t) \sigma^x_i \rangle = \cos(2\lambda t)^2 + \sin(2\lambda t)^2 = 1$. Starting in the $h_c = 0$ limit and moving toward high transverse magnetic field $h_0 >> \lambda L$, we numerically integrate Eq.\ref{eq:star_exact} and study the autocorrelation function on site$-i$ (Fig.\ref{fig:star_analytic}). In the low field limit, we expect exponentially small modifications to pure cosine oscillations as we have recreated from ~\cite{lucas2019quantum}, and in the high field limit, $\sigma^x_i$ should completely break down as $\sigma^z_c$ decoheres and operator weight spreads to all sites rapidly. As $\sigma^z_c \rightarrow [\sigma^y_0, \sigma^x_0]$, operator weight can be distributed to all sites $j \neq i$, and the rapid rotation of operators on the c-qubit aliases away coherent growth on site-i.

In Fig.\ref{fig:star_analytic}(a) we see that coherent oscillations are apparent for $h/\lambda L ~ \mathcal{O}(1)$ out to $t\lambda \sim 30$ and slow decay with increasing $h$. Above $\log[h/\lambda L] > 0$, oscillations are barely visible and dynamics are dominated by exponential decay with no revival even out to time $t\lambda \in [0, 200]$. This is more clearly depicted in Fig.\ref{fig:star_analytic}(b), where we take the long-time average of the autocorrelation function:
\begin{equation}
    A(i,t_0) = \int_{t>0}^{t_0} |\langle \sigma^x_i(t)\sigma^x_i\rangle | dt.
\end{equation}
The autocorrelation function transitions sharply at $h/L \sim \mathcal{O}(10^{-1})$, where $\sigma^x_i$ no longer has significant weight on the initial site. For small $h$, the autocorrelation function is a nearly pure $\cos{2\lambda t}$ and decay that grows with $h$. Near the transition, $A(i,t_0)$ sharply decays to $\sim 1/L$ for $h/\lambda L > \mathcal{O}(10^{-1})$. We provide a Fourier analysis of Fig.\ref{fig:star_analytic}(a) in (c), where the transition is more clearly resolved. The oscillatory part ($\epsilon_0$) decreases (elongates in time) above $h/\lambda L = 1$ with an exponential decay rate that peaks at the transition point. Above the transition point, the dynamics are no longer captured by a decaying cosine function, but crossover to exponentially damped operator dynamics. The large magnetic field on the central site initially prevents operator weight from leaving site$-i$ as shown by the elongating decay in (a). At this point the magnetic field is dominating the operator dynamics and rapidly rotates $\sigma^z_c$ into eventual two-body operators that live in a superposition on all sites. In this regime, operator weight on the initial site saturates with a scaling like $1/L$. 

Adding a mix of on-site frustrating magnetic fields is not sufficient to fully scramble information. Applying fields along the $x, z$-direction, local operators do not spread throughout full operator Hilbert space as the system can still be described as a collective $S = L/2$-spin interacting with a qubit. Similarly the steady-state entanglement entropy remains independent of $\lambda$ and system size, while the growth rate is determined by $\text{min}[1/\lambda, 1/h_c]$ and does not slow when when deep into the coherent regime $\lambda L > h_c$ [see Supplemental for greater details].

\section{Numerical Results}
After discussing the nearest neighbor mixed-Ising chain and the star Ising model, we now seek to understand the dynamical behavior in the full, ring-star system. Here we study the dynamics of the star-local model using exact diagonalization for systems up to $L+1 = 13$ and Krylov subspace expansion techniques for evaluating the Schrodinger ODE for sizes up to $L+1 = 22$. We employ periodic boundary conditions $i= L = 0$. In the previous section we revealed that a dynamical operator growth transition occurs as a function of $h_c/\lambda$ on the Ising star graph that allows for rapid operator growth from the central qubit or coherently protected operator dynamics on the central qubit. When the operator dynamics are coherently oscillating on the central site, this leads to slow two-time correlator growth. Here we investigate the fate of this transition, its effect on the secondary, local Ising channel for quantum scrambling, and whether fast-scrambling is achievable in simple local-nonlocal construction. We know that the mixed-field Ising model is nonintegrable and capable of scrambling information throughout the full operator Hilbert space, but does a simple nonlocal qubit rapidly enhance this process?

Firstly, we calculate the average adjacency level ratio $\langle \bar{r} \rangle $ of the full Hamiltonian Eq. \ref{eq:star_local_Ham} [see Appendix Fig.\ref{fig:energy_stats}]. We solve the full spectrum exactly considering the parity conserving and $k = 0$ sector of the Ising chain $(L=15)$ and confirm that for the ring-star model and nonintegrable Ising model that all points in phase space we consider indeed follow GOE random matrix statistics with $\langle \bar{r} \rangle \approx 0.53$. Though the level spacing provides a first check of nonintegrable dynamics, it is not sufficient in capturing strongly coherent effects inherent to fractionalized regions of the Hilbert space or rare states that exhibit confinement or slow growth~\cite{Rakovsky2019}.

\begin{figure}
    \centering
    \includegraphics[width=0.5\textwidth]{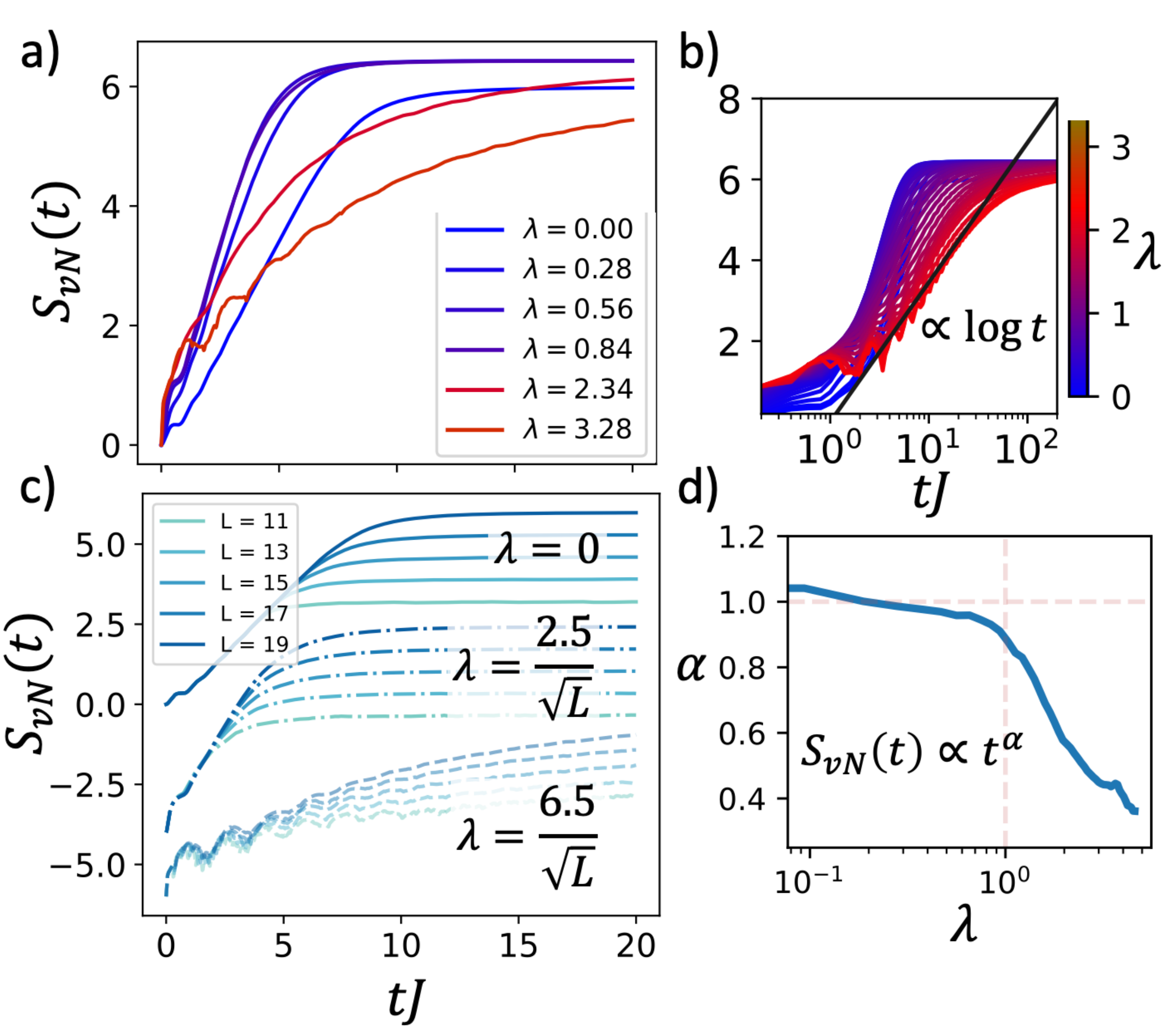}
    \caption{\textit{Entanglement spreading in the ring-star model: } (a) $S_{vN}(t)$ as a function of $\lambda$, (b) plotted on a semilog axis in time ($tJ$), and (c) $\lambda$ reparameterized to $\lambda/\sqrt{L}$, system size $L$. 
    (d) Polynomial fit of the intermediate time entanglement growth behavior $S_{vN} \propto t^\alpha$. 
    System size $L+1 = 22$ and $J, h, g = [1.0, 1.05, 0.45]$.}
    \label{fig:ring-star_rt}
\end{figure}

\subsection{Entanglement Growth} The first test as to the general information scrambling capacity of this model is to understand the entanglement entropy dynamics. We numerically investigate entanglement growth $S_{vN}(t)$ as a function of c-qubit coupling. To understand the infinite temperature information dynamics of the system, we work with a product state with energy density equivalent to that of an infinite temperature state $\langle H_i \rangle = 0$. As the system exhibits GOE random matrix statistics for all parameters of the Hamiltonian tuned here, we expect that any such effective infinite temperature pure state will indeed obey ETH. The state we work with here is the polarized $|+y\rangle$ spin-state, which has been used to exemplify high-energy quench dynamics of the mixed-field Ising chain previously~\cite{Belyansky2020}.

In the limit $\lambda = 0$, we have the local, nonintegrable Ising model which exhibits ballistic entanglement growth for times up to $L/v_k$, where $v_k$ is the maximal dispersion associated with quasiparticle momentum $k$. This timescale is upper-bounded by $L / (2J)$ which provides the characteristic timescale of nearest-neighbor interactions multiplied by the length of the system. The factor of two comes from periodic boundary conditions and simply captures the longest path between spins. In the $\lambda = 0 $ curve in Fig.\ref{fig:ring-star_rt}(c) we see that for increasing system size, entanglement saturation occurs on a timescale that scales linearly with $L$ and the maximal quasiparticle velocity remains independent of system size. Entanglement saturates like $L\log(2)$ as the effective infinite temperature state explores the full Hilbert space of the system. With the structure of correlations reaching the scale of the Hilbert space $O(2^L)$ we then expect entanglement to be similar to the log of full operator space complexity.

As we slowly increase the nonlocal coupling to the central c-qubit, we expect that nonintegrability is maintained and now the shortest path between sites is next-nearest neighbor, mediated by the central qubit. The timescale for the c-qubit mediated entanglement spread is $2/\lambda$, where the factor of $2$ comes from the second order interaction with the central c-qubit that allows correlations to develop between sites $i,j \neq 0$. In the extremely weak regime $\lambda << J, h, g$, the c-qubit qubit can similarly be thought of as a cavity, where the finite size Hilbert space can be disregarded.

Including the c-qubit modifies the early-time entanglement growth, as operators grow ballistically due to local transport and super-ballistically with an additional nearest-neighbor c-qubit mediated contribution. We see in Fig.\ref{fig:ring-star_rt}(a) that for $L+1 = 20$ as $\lambda \rightarrow h$ the linear coefficient of entanglement growth grows continuously and the saturation value is modified by the addition of a single qubit. In this enhanced rapid scrambling regime, if we reparameterize the central qubit coupling $\lambda \rightarrow \lambda/\sqrt{L}$ such that the effective spin-spin interaction is not extensive $J_\text{eff} \sim \lambda^2.0 / L$, we find that the coefficient for linear growth is roughly independent of system size (Fig.\ref{fig:ring-star_rt}(b) $\lambda  = \frac{2.5}{\sqrt{L}}$). Entanglement entropy growth rate that increases with system size is indicative of fast-scrambling behavior~\cite{Belyansky2020}.

For $\lambda >  h$, we see a surprising entanglement transition. The early time growth defined as $t < 2\frac{1}{\lambda}$ exhibits rapid entanglement growth but becomes sub-ballistic at intermediate times $2\frac{1}{\lambda} < t < t_{sat}$. In Fig.\ref{fig:ring-star_rt}(b) the width of this intermediate timescale grows exponentially with increasing $\lambda$ as exhibited by the logarithmic scale in time, while $S_{vN}(t\rightarrow\infty)$ remains largely unchanged. We perform a polynomial fit of the entanglement growth in Fig.\ref{fig:ring-star_rt}(d) over the region $\log[2] < S_{vN}(t) < S_{vN; sat}$ and find that in the local and fast-scrambling regime, entanglement continues to grow ballistically with exponent $\alpha \sim 1.0-1.1$. For $\lambda > \lambda_c$, $\alpha$ monotonically decreases with $\lambda$ as expected if $S_{vN} \propto \frac{1}{\lambda}\log[t]$.

\begin{figure}
    \centering
    \includegraphics[width=0.35\textwidth]{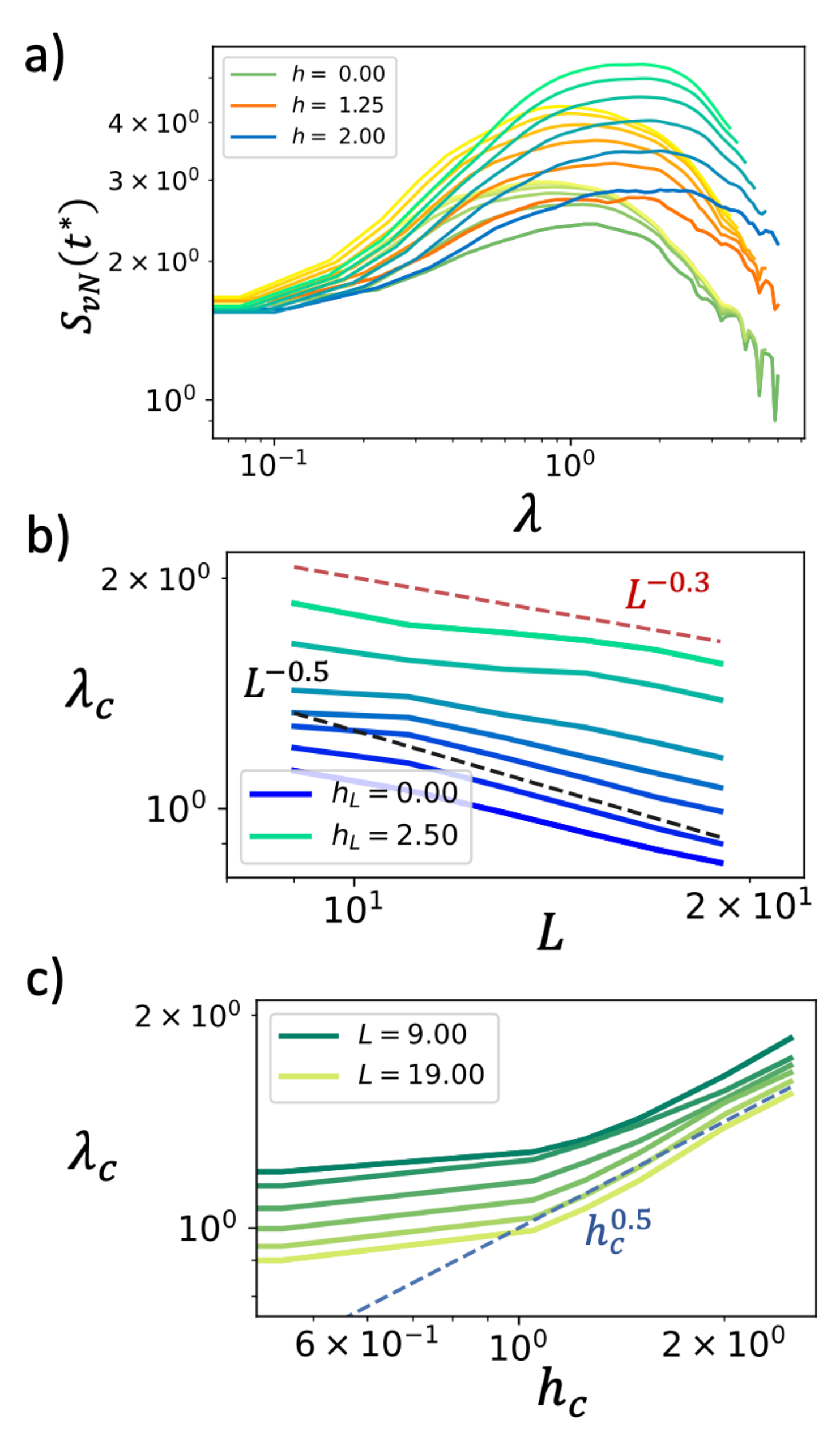}
    \caption{\textit{Identifying entanglement dynamics crossover:} (a) Entanglement entropy calculated at time $t^\star \sim 3.5tJ$ plotted as a function of c-qubit coupling $\lambda$ ($S_{vN}(t^\star) / t^\star \sim v_b$) for three transverse magnetic fields and various system sizes ($h_c \in [1.05, 2.5]; L \in [11-21,2]$.  For nonzero $\lambda$, entanglement growth rate varies as a function of $h_c, L$. Crossover point $\lambda_c$ determined as maxima of the curves in (a) with $\lambda_c(L)$ and $\lambda_c(h_c)$ plotted in (b,c), respectively. In the respective $h_c$ dominant and $L$ dominant regimes, the critical point find $\lambda_c \propto L^\gamma h_c^\kappa$ with $\gamma = -0.52\pm 0.02 (1)$ and $\kappa = 0.52\pm 0.05(1)$.}
    \label{fig:ring-star_PT}
\end{figure}

In order to determine the location of the phase transition and relevant scalings, we analyze the entanglement entropy at a fixed time $t^*$. We choose a time such that the $S_{vN}(t^*, \lambda = 0)$ is roughly $\frac{1}{2}S_{vN}(t\rightarrow\infty)$. In Fig.\ref{fig:ring-star_PT}(a) we plot $S_{vN}(t^*)$ as a function of two magnetic fields $h$ depicted by solid/dashed curves and as a function of system size $L$. We extract the maxima of  Fig.\ref{fig:ring-star_PT}(a) and perform a size and field scaling and find that the fast-to-slow scrambling transition value occurs like $\lambda_c \propto L^{(-0.52 \pm 0.05)}h_c^{(0.52\pm 0.02)}$, (b,c) respectively. In the high $h_c$ limit, the transition becomes nearly independent of system size ($h_c = 2.5$ curve in Fig.\ref{fig:ring-star_PT}(c)) and in the low field limit, is dominated by system size: $h_c = 0.05-0.75$ curves in Fig.\ref{fig:ring-star_PT}(c). Greater details on identifying the entanglement transition are included in the Supplemental Information. This contrasts the dynamical transition observed in the star Ising model with a transition that occurs at $\lambda_c = h_c/L$.

The novel entanglement transition mediated by the c-qubit dynamics is extremely surprising in that not only does the nonlocal coupling provide a secondary channel for distributing entanglement, but in the strong coupling regime it inhibits growth of even the local Ising interactions with which it commutes. The problem is similarly interesting in that it is completely disorder-free, so the slow information growth in the system can be attributed to purely coherent effects. In order to shed more light on how this central coupling serves to rapidly/slowly scramble quantum information we examine the transverse and longitudinal OTOCs ($C_{xx}, C_{zz}$).

\begin{figure}
    \centering
    \includegraphics[width=0.485\textwidth]{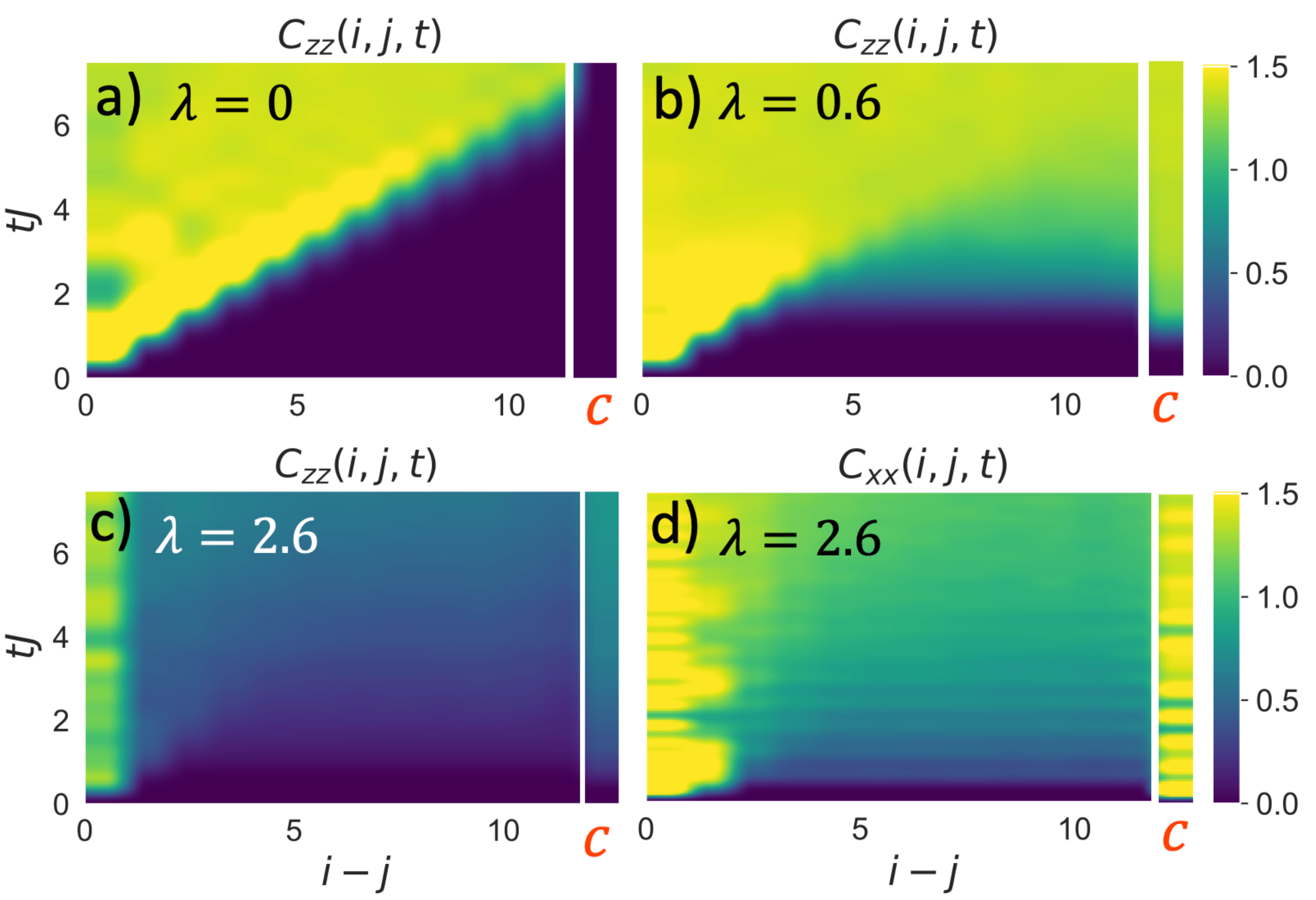}
    \caption{\textit{Space-time OTOC spreading: } (a-c) OTOC for longitudinal spin component $C_{zz}$ for $\lambda = {0, 0.6, 2.6}$, respectively. Initial linear light cone spreads on top of c-qubit mediated operator weight that spreads super-ballistically on all sites after an initial wait time that scales with $min[1/h_x, 1 / \lambda]$. For $(\lambda/\sqrt{L})/h \geq 1$, $C_{zz}$ is suppressed on all sites beyond a weak initial light cone. (d) In contrast, for $\lambda =2.6$, transverse OTOC $C_{xx}$ rapidly fluctuates on the initial site-$i$ and spreads to nearest neighbors $i+1$ and the central site $L$ on the order of $t = 1/\lambda$. The light cone profile is no longer visible and $C_{xx}$ rapidly oscillates with frequency $2\lambda$ for sites $|i-j| > 1$. System size $L+1 = 14$ and $J = 1.0, h = 1.05, g=0.45$.}
    \label{fig:otocDensity}
\end{figure}

\subsection{Operator Spreading}
Starting from an initial state selected from the Haar measure, as to approximate infinite temperature state, we examine how operators initially prepared on site $i=0$ spread under the influence of Ising interactions, where site $c$ represents the central qubit. When calculating the OTOC we simplify Eq.\ref{eq:OTOC_full}, as the local $\hat{X}, \hat{Z}$ operators are Hermitian and initially commute on all sites at $t = 0$. We then explicitly evaluate
\begin{equation}
    C_{VW} = 1 - \langle \hat{W}(j,t)\hat{V}(i,0)\hat{W}(j,t)\hat{V}(i,0)\rangle
\end{equation}
In Fig.\ref{fig:otocDensity}(a), when considering only local interactions we see that operator weight develops according to a ballistic light cone with a bubble-like profile that saturates on all sites following the wavefront: $C_{zz}(r,t) \propto \frac{t^{2r}}{r!^2}$. With small, nonzero $\lambda$ (b) the light cone remains apparent but now spreads on top of growing operator weight distributed by the central qubit. The central qubit super-ballistically distributes operator weight to all sites on a timescale like $\frac{2}{\lambda}$. The growth timescale on the c-qubit is half that of the bulk spins, as the $\sigma_z$ operator must decohere on the initial site and again on the central site and hence no longer commutes with the Ising interaction. For $\lambda << h_c$, $1/\lambda$ sets the timescale for this process. In the strong coupling regime (c), a weak light cone remains but transfers a small fraction of the original operator weight.
Operator spreading becomes increasingly restricted where the lightcone profile is only visible for few $tJ$ and $C_zz(j,t)$ grows increasingly slowly on sites far from the initialized operator. This behavior is puzzling; the highly nonlocal coupling c-qubit with increasing interaction leads to unintuitively slow local operator spreading.

Here we can draw insight from the analytic results on the star-Ising model (Fig.\ref{fig:star_analytic}). The $L-$body interaction on the central qubit acts to coherently project operators into $\sigma^z$, while rapidly aliasing orthogonal operators. Once $\sigma^z$ grows on the central qubit, it develops a coherent lifetime with operator weight decaying like $\sim e^{-1/\lambda}$. $C_{zz}$ characterizes how operators no longer commute with $\sigma^z$, and in the same vain as the star-Ising model, we expect operators on the central site to be strongly projected into the $z-$subspace and coherently oscillate like $2\lambda$. As the central c-qubit is highly/fully entangled with the remaining chain, the coherent projection on the central site then similarly restricts how rapidly the many-body state of the $L-$spins decoheres from the $z-$subspace. We expect $C_{zz}(t)$ to exhibit slow behavior as $z-$operators are strongly driven on all sites. We then examine $C_{xx}(t)$, which captures the rapid growth of $\hat{Z}$. In Fig.\ref{fig:otocDensity}(d) we see that $C_{xx}(t)$ continues to grow like $\mathcal{O}(1)$ across all sites, with strong, oscillatory behavior on the central qubit ($i-j = c$). $C_{xx}(j,t)$ on all sites similarly oscillates at frequency $2\lambda$ but shifted by half a period compared to $C_{xx}(c,t)$.

\begin{figure}
    \centering
    \includegraphics[width=0.485\textwidth]{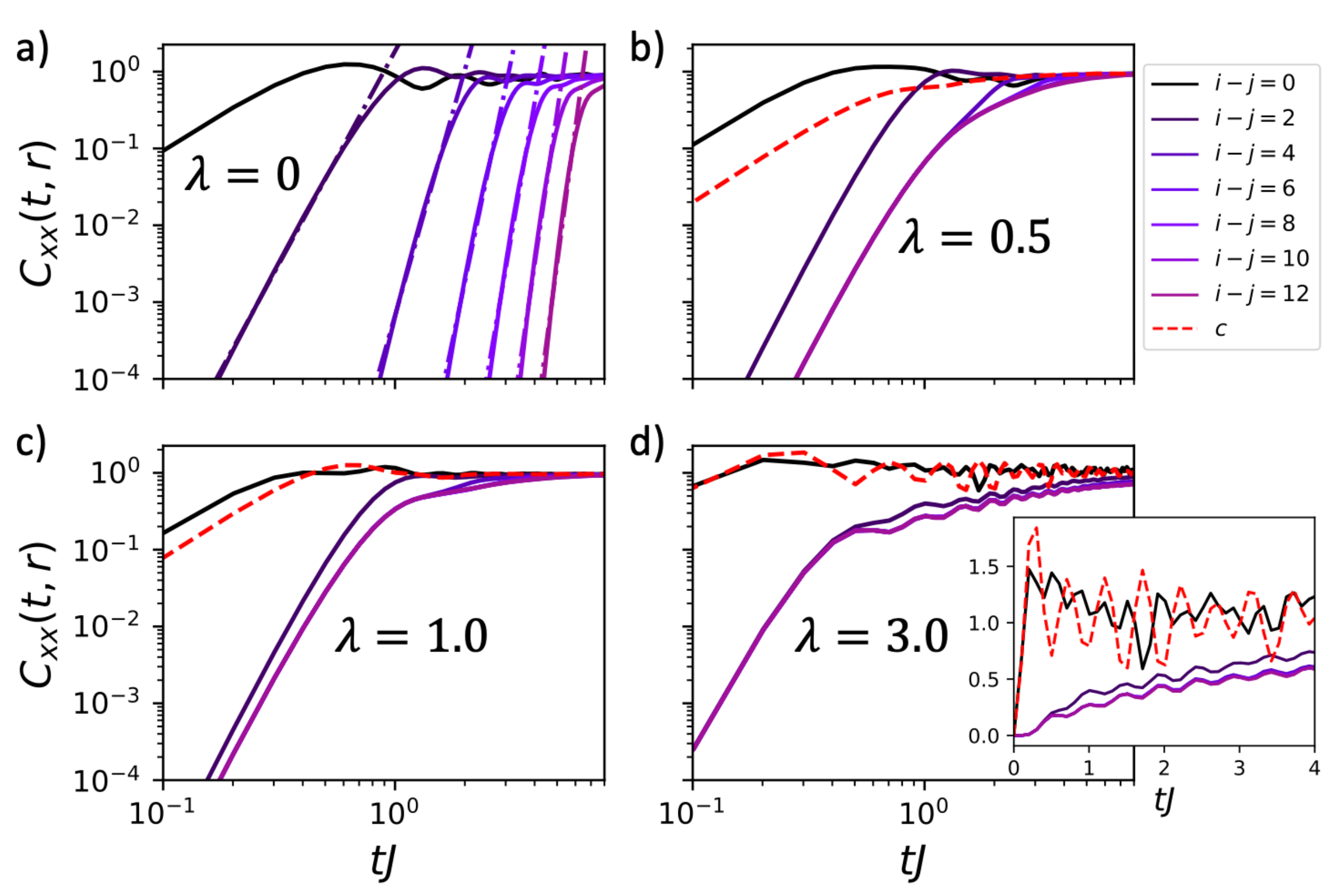}
    \caption{\textit{$C_{xx}(|i-j|, t)$ spreading in the ring-star Ising model:} Real-time evolution of $C_{xx}(i-j, t)$ for varying c-qubit coupling $\lambda$, plotted on a log-log scale. $\lambda = 0$ exhibits lightcone spreading $C_{xx} \propto \frac{t^{2(i-j)}}{(i-j)!^2}$. In the fast-scrambling regime ($\lambda< \lambda_c$) (b,c), operator weight grows homogeneously across all sites at half the rate as the on the c-qubit site ($C_{xx}(c,t)$, blue). (b,c). (d) As $\lambda \sim \mathcal{O}(1)$, $C_{xx}(c,t)$ becomes $\sim \mathcal{O}(1)$ at $t \propto \lambda^{-1}$. Once $C_{xx}(c, t)$ becomes $\mathcal{O}(1)$, the rapid growth on spin sites $|i-j| > 1$ slows and coherent oscillations develop. For $\lambda = 3.0$ and focusing on distant sites (purple) and $tJ\in [1, 5]$, the operator growth after c-qubit saturation becomes slower than that observed for $\lambda = 1.0$. System size $L+1 = 15$ and $J = 1.0, h = 1.05, g=0.45$.}
    \label{fig:xx_med}
\end{figure}

On the initial site $i$, as $\sigma^z_i$ decoheres under the transverse magnetic field it no longer commutes with itself nor the Ising interaction. This leads to a nonzero $C_{zz}(t)$ on site$-i$ and subsequent operator weight growing on nearest neighbors and the central qubit. Once operator weight grows on the c-qubit, the associated decoherence time of $\sigma^z_c$ is exponential in $\lambda$ and as weak operator weight leaks onto bulk spins, the same coherent oscillations and slow decay is imparted on the local spin-chain. We gain further insight by examining the individual curves that make up the density plot in Fig.\ref{fig:otocDensity} as Fig.\ref{fig:xx_med}. We see the polynomial growth associated with the light-cone spreading on top of the exponential c-qubit mediated growth (a-c).In the slow scrambling regime (Fig.\ref{fig:xx_med}(d)), we find that for $\lambda = 3.0$ that $C_{xx}(c,t)$ becomes $\mathcal{O}(1)$ at $t_c = \pi/2\lambda$ (dashed, red) and after time $t_\lambda = \pi/\lambda$ the rate of change of $C_{xx}(i-j,t)$ decreases dramatically (solid). Once $\sigma^z$ exists on the central qubit, orthogonal operators on the c-qubit and bulk sites rapidly fluctuate and lead to an effective wait time, slowing the amount of operator growth under the action of the Ising$-zz$ interaction.

\begin{figure}
    \centering
    \includegraphics[width=0.475\textwidth]{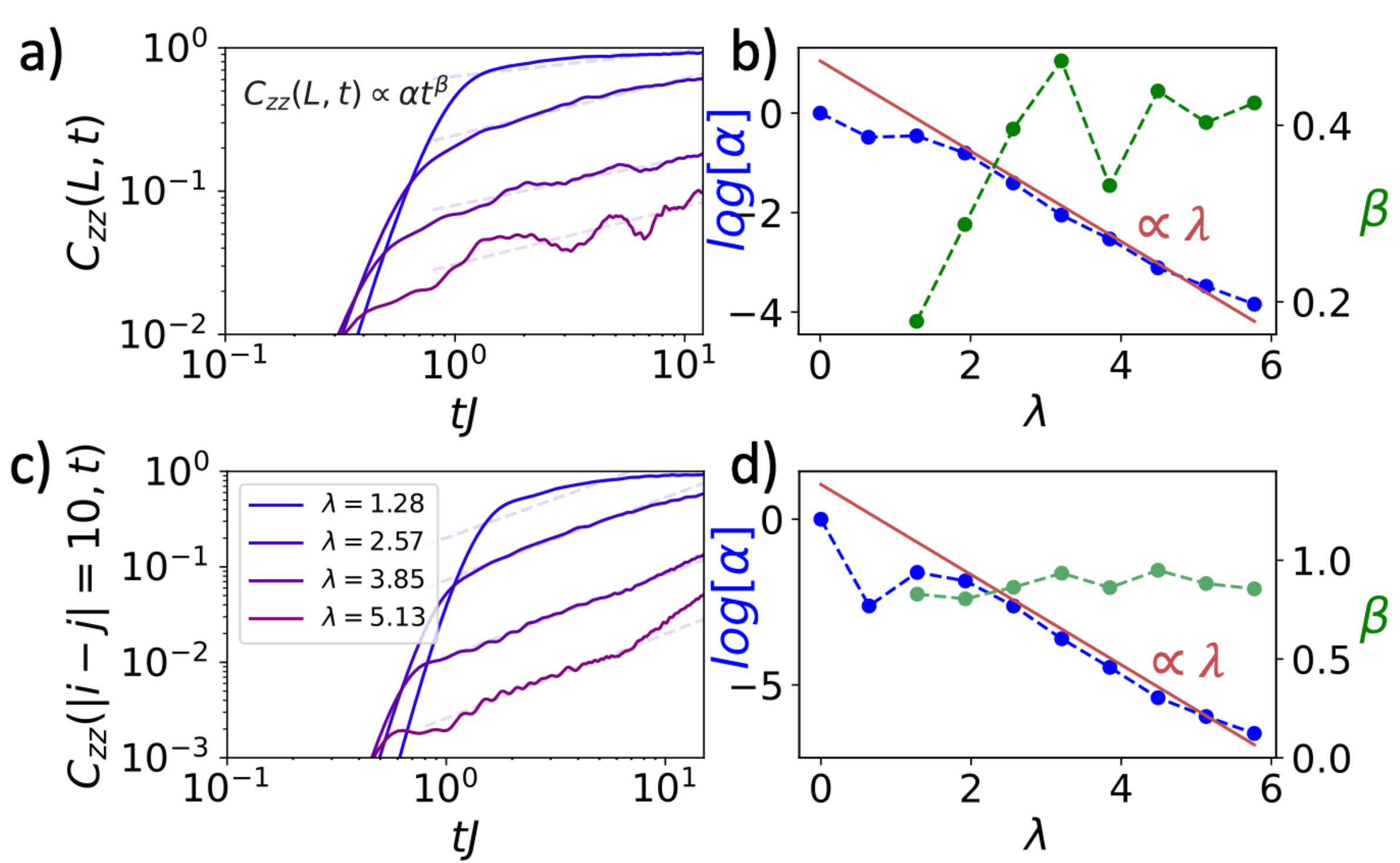}
    \caption{\textit{$C_{zz}(t)$ spreading in the ring-star Ising model:} Longitudinal OTOC $C_{zz}(t)$ on (a) site $j = 10$ and (c) c-qubit $j = c$ with real-time evolution plotted on a log-log scale. (b, d) Polynomial fit at intermediate times $C_{zz}(t) \propto \alpha t^\beta$. As $\lambda$ increases, coefficient of longitudinal operator growth becomes exponentially suppressed (top to bottom curve on log scale) $\log{\alpha} \propto \lambda / \sqrt{L}h^2$. Operator growth on distant spin sites is approximately linear $\beta \sim 1$ in the slow scrambling regime and coefficient $\alpha = e^{-\frac{3}{2}\frac{\lambda}{\sqrt{L}h^2}}$. (d) $C_{zz}$ the c-qubit site grows with exponent $\beta_c \sim 0.5$ and coefficient $\alpha_c = e^{-\frac{\lambda}{h^2}}$. Operator growth on bulk spin sites is expected to behave like $C_{zz}(j, t) = C_{zz}(L, t)^2$, as the effective c-qubit mediated spin-spin interaction is second order process in $H_{\lambda}$.}
    \label{fig:slow_growth}
\end{figure}

A complementary perspective for operators orthogonal to $\hat{Z}$ is observed by $C_{zz}$ (Fig.\ref{fig:slow_growth}. Looking at the OTOC in the slow scrambling regime, operator weight on the c-qubit and bulk sites grows like $h^2t^6$, as captured by early-time expansion, up to time $t_\lambda = \pi/2\lambda$ [see Supplemental for greater details]. For $t > t_\lambda$, operators become essentially projected onto $z-$operator subspace due to the coherent lifetime of operators on the central qubit. Fitting the OTOC to $\alpha t^\beta$ shows that operator weight decoheres on the c-qubit with exponent $\beta_c \sim 0.45$, while on bulk sites it is nearly linear with $\beta_j = \sim0.85$. On all sites, the coefficient of growth becomes exponentially suppressed in $\lambda$ with $\log[\alpha] \propto -\lambda$. This can be understood heuristically: once operator weight lies on the central site, it coherently oscillates between sites like $2\lambda$ and leads to a waiting time during which little to no operator weight has decohered from the $z-$subspace as it oscillates between $z-I$. The amount of operator weight that decoheres in-between these waiting periods is $\int_{t_{c1}}^{t_{c2}} e^{-\lambda}$ so the rate of operator growth into this orthogonal subspace is independent of time and leads generically to $C_{xx}(j!=c, t) \propto e^{-\lambda}t$. This slow growth for simple two-body body operators outside of the $z-$subspace continues to be exponentially suppressed for greater complexity many-body operators. The sublinear growth of OTOCs into the bulk of operator Hilbert space then generically provides $\log[t]$ entanglement entropy growth, as we have observed for quantum quenches from the $|+y\rangle$ state.

\section{Discussion}

Here we have shown how the collective interactions between many constituent spins and a central auxiliary qubit is both able to rapidly scramble information across all degrees of freedom and restrict state/operator growth from exploring the full Hilbert space. From an entanglement entropy picture, the central bit entangles rapidly with its surrounding environment, the state space of the qubit simply being $q=2$. Treating it as a 2-level system in an effective bath, the extensive bath interactions and on-site magnetic field frustrate the central qubit and lead to a rapidly fluctuating spin-moment when $N\lambda \sim h_c$. In the operator picture this leads to rapidly decohering operators that quickly lie in a superposition of states $\{\hat{X}, \hat{Y}, \hat{I}, \hat{Z}\}$. For $\lambda >> h_c$, $\{\hat{Z}\}$ operators are strongly driven to the qubit and therefore do not fully decohere while $\{\hat{X}, \hat{Y}\}$ become essentially echoed-out. 

This picture is more clearly developed when we again turn-off local interactions, $J=0$. In this case only the central bit is able to mediate entanglement throughout the system. As a function of $\lambda/h_c$ the half-chain entanglement entropy grows with a velocity like $1/\lambda$ until $\lambda ~ h_c$ and surprisingly saturates with an entropy that is independent of system size. The entropy growth rate and saturation value do not change when crossing over into the strong-coupling regime. Though the half-system entropy value is identical, we know from an operator picture that scrambling occurs slowly on the leaves of the star graph, leaving the central qubit to scramble rapidly while other sites scramble slowly. From both the operator and entropy picture, we know that the central bit is highly entangled with the system and the saturation entropy remains fixed regardless of coupling; therefore once the central bit becomes highly entangled, it similarly cannot scramble quantum information throughout. The saturation in information capacity of the star-graph remains fixed as a function of $\lambda$, so by tuning $\lambda$ we tune whether entanglement is frozen in a highly entangled auxiliary bit or distributed globally. Understanding the information dynamics of the star-graph then informs the surprising behavior observed in the fully interacting star-local Ising system. Once the central qubit becomes highly entangled with it's environment, it similarly acts to project operators into the $z-$subspace, inhibiting the local, nonintegrable Ising interactions from fully scrambling the system and reaching full thermalization quickly.

As the commutivity graph that translates operators from site$-i$ to site$-j$ requires operation of the Ising$-zz$ interactions, which similarly commutes with the central spin coupling, local operator growth spreading is inhibited on the same decoherence timescale. The local channel is effectively turned off by the slow operator decoherence. If the local interactions were orthogonal to the central coupling, we expect different operator spreading results. Here we take a simple extension of our model and change the form of the auxiliary coupling to be Ising$-xx$ which we term a compass, ring-star Ising model. Now the the auxiliary coupling projects the system into $\hat{X}$ subspace, which no longer commutes with the local $-zz$ interactions. This should allow many-body operators to propagate throughout but inhibit the growth of operator strings containing $\hat{Y}, \hat{Z}$. Translating the operator picture to the entropy picture, we are allowing higher order many-body operators to develop through local channels, so the intermediate time saturation value should be significantly larger and scale extensively with system size, the logarithmically slow growth of orthogonal operator strings should then extend to long times as operators can decohere to larger regions of the Hilbert space.

\begin{figure}
    \centering
    \includegraphics[width=0.45\textwidth]{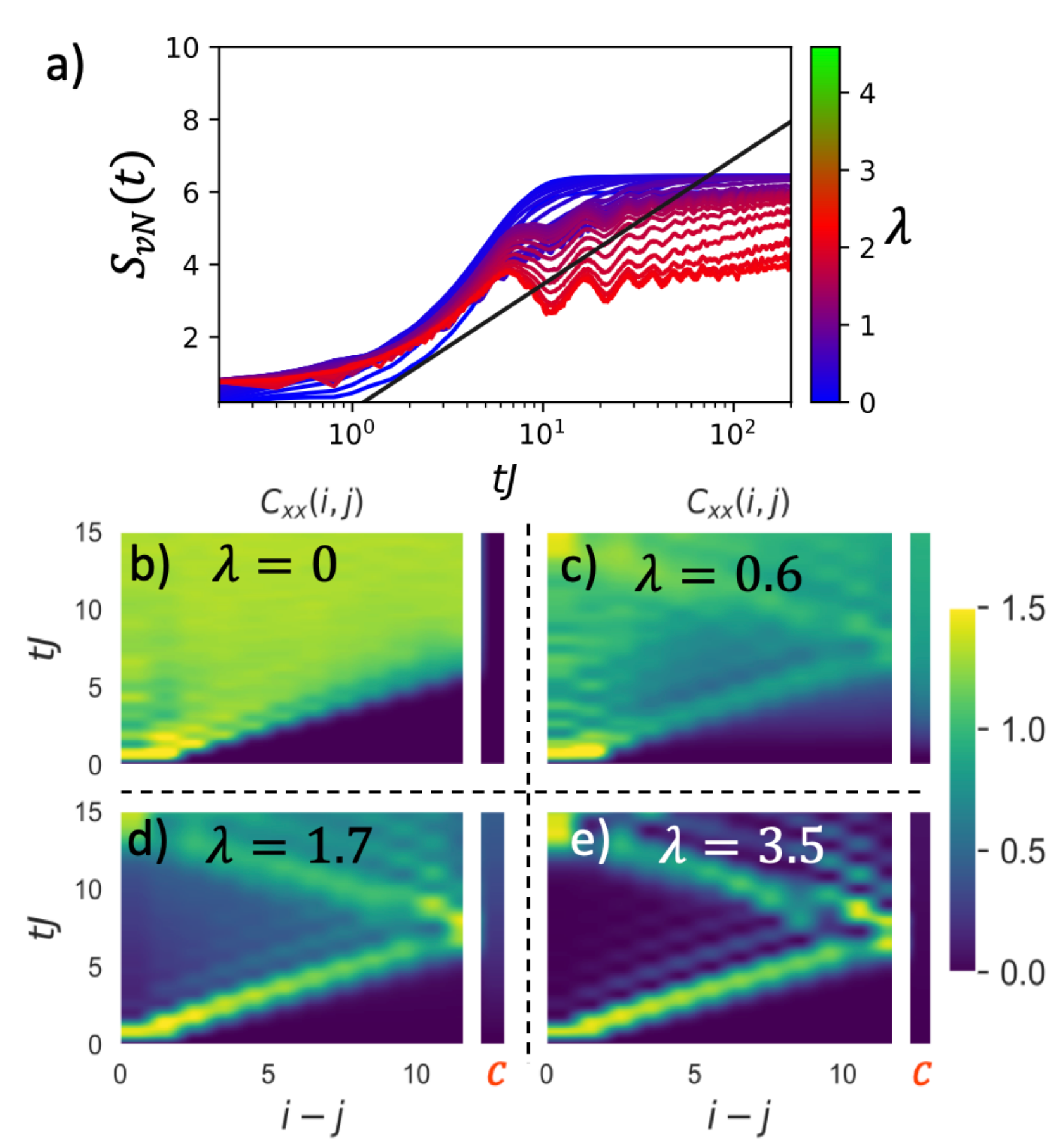}
    \caption{\textit{Information spreading in the transverse ring-star model ($\lambda \sum_i^{L-1} \sigma^x_i \sigma^x_c, J=1.0$): }(a) Early-time entanglement entropy growth crosses over from ballistic to sub-ballistic with increasing $\lambda$, plotted on a semi-log scale. Entropy depicts a prethermal phase with entanglement that saturates around $t$, then exhibits characteristic MBL, logarithmic growth $\propto \log(t)$ out to late times $tJ = 200$. Early-time entropy growth increases smoothly with $\lambda$ until reaching a discontinuity at $\lambda \sim 1$. (c, d, e, f) OTOC for transverse spin component for $\lambda = {0, 0.84, 2.0, 3.0}$, respectively. In contrast to the parallel ring-star Ising model, the light cone becomes increasingly visible as $\lambda$ increases. (c) In the nonintegrable, purely local case, operator scrambling saturates following the light cone. As c-qubit coupling grows, operators orthogonal to $\sigma^x$ rapidly fluctuate, so that following the light cone, $\sigma^x$ operator weight dominates each site. (e) The light cone boundary eventually becomes the only region where operators are orthogonal to the c-qubit coupling and propagate under local Ising interactions. System size $L = 13$ and $J = 1.0, h = 1.05, g=0.45$.}
    \label{fig:xxCouple}
\end{figure}

In Fig. \ref{fig:xxCouple} we observe exactly this behavior. Noting that the magnitude of the orthogonal field on the c-qubit changes from $h \rightarrow g = 0.45$, the inhibited operator growth goes as $\lambda_x / g^2$. The entanglement entropy (a) following a quench from $|+y\rangle$ transitions from local spreading, fast-spreading, to confined for $\lambda_x > 0.5$. For times $tJ < 10$, the system exhibits ballistic operator growth regardless of coupling but reaches an intermediate saturation value $S_{vN} \approx 3$ with slow growth extending out to late times. The OTOC agrees identically, where we have the same ballistic spreading lightcone with fully scrambled operators following the wavefront (b). With increasing coupling (c) we observe a small amount of operator weight that spreads super-ballistically across all sites, a well defined wave-front, and a suppression in the OTOC behind the wavefront. This becomes more dramatic deeper into the confined regime (d,e), where the wavefront becomes the only region where operators orthogonal to $\hat{X}$ may be found. As the local Ising interaction is orthogonal to the operator subspace in which the c-qubit protects, the wavefront may freely propagate but operators behind the wavefront are aliased to predominantly span strings of $\hat{X}$ and $\hat{I}$ operators.

\section{Conclusion}

We have interrogated the quantum entanglement and operator dynamics in a central spin-bath design. The local Ising chain acts as a structured, thermalizing bath that when coupled to a nonlocal central qubit, is able to rapidly scramble quantum information across the system up until the information capacity of the c-qubit is quenched. When the central bit is rapidly quenched in the strong coupling regime, it leads to operator confinement in sectors of the Hilbert space spanned by $\hat{I}$ and operators parallel to the central qubit coupling. Entanglement entropy grows slowly out to late times as observed in effective infinite temperature states.

This model admits unique limits; where, for weak coupling it exhibits super-ballistic OTOC spreading and fast-scrambling as supported by previous works examining a similar star-graph structure, and a confined quantum Zeno-like limit in which the complexity of operator growth is inhibited by extensive, coherent interaction with the two level central mode. This confined regime mirrors a larger class of models which admit prethermal or localization physics: disorder driven many-body localization, projective or weak quantum measurement, floquet-driven prethermalization. Our star-local work is similar in that the c-qubit drives operator fluctuations on the leaves on a timescale like $1/\lambda$ once $\lambda$ is greater than noncommutting fields on the central site. In contrast, the qubit retains its infinite-range information sharing capacity compared to the external drive scenario in floquet physics. In the measurement theory sense, once the central qubit becomes nearly maximally entangled with the Ising spin chain bath, its as if the bath projects the central spin which in turn freezes the information sharing dynamics with which the bath is entangled. And finally in the MBL case, instead of local integrals of motion that are nearly integrable with exponentially small overlap, here we have global operators orthogonal to the auxiliary coupling that have a logarthimically slow thermalization timescale.

This work provides a novel mechanism for exploring a host of quantum information dynamics. Here we outline the existence of a dynamical transition that occurs under static Hamiltonian dynamics, and in future work it would be fruitful to investigate the nature of the transition and how similar physics can be observed in RUCs. Tracing out the central qubit to produce non-Hermitian physics may similarly illuminate how this model relates to strong periodic driving prethermalization. As the inverse problem, it would be interesting if exotic non-Hermitian physics has a hidden unitary quantum system analog, where multiple ancillary qubit degrees of freedom produce the observed non-Hermitian dynamics. Future work should also seek to understand how this transition persists with larger c-qudit towards an infinite bosonic fields or multiple dissipative modes. It would be an interesting engineering application if uniform coupling to a bosonic mode or central-qubit can protect against dephasing errors in the environmental qubit platform.

\section{Acknowledgements}
J.C.S and N.T. would like to thank Sumilan Banerjee, Chandrasekhar Ramanathan, Brian Skinner, Xiaozhou Feng, Shi Feng, and Sayantan Roy for useful discussions. This material is based upon work supported by the U.S. Department of Energy, Office of Science, Office of Basic Energy Sciences under Award Number DE-FG02-07ER46423. Computations were done using the QuSpin python package ~\cite{quspin} on the Unity cluster at the Ohio State University.

\newpage
\bibliography{references}
\newpage
\onecolumngrid

\begin{center}
\textbf{\large Supplemental Material for "Entanglement Dynamics between Ising Spins and a Central Ancilla"}
\end{center}

\setcounter{section}{0}
\renewcommand{\thefigure}{S\arabic{figure}}
\setcounter{figure}{0}

\section{Operator Spreading on the Star Graph}

\begin{figure}
    \centering
    \includegraphics[width=0.65\textwidth]{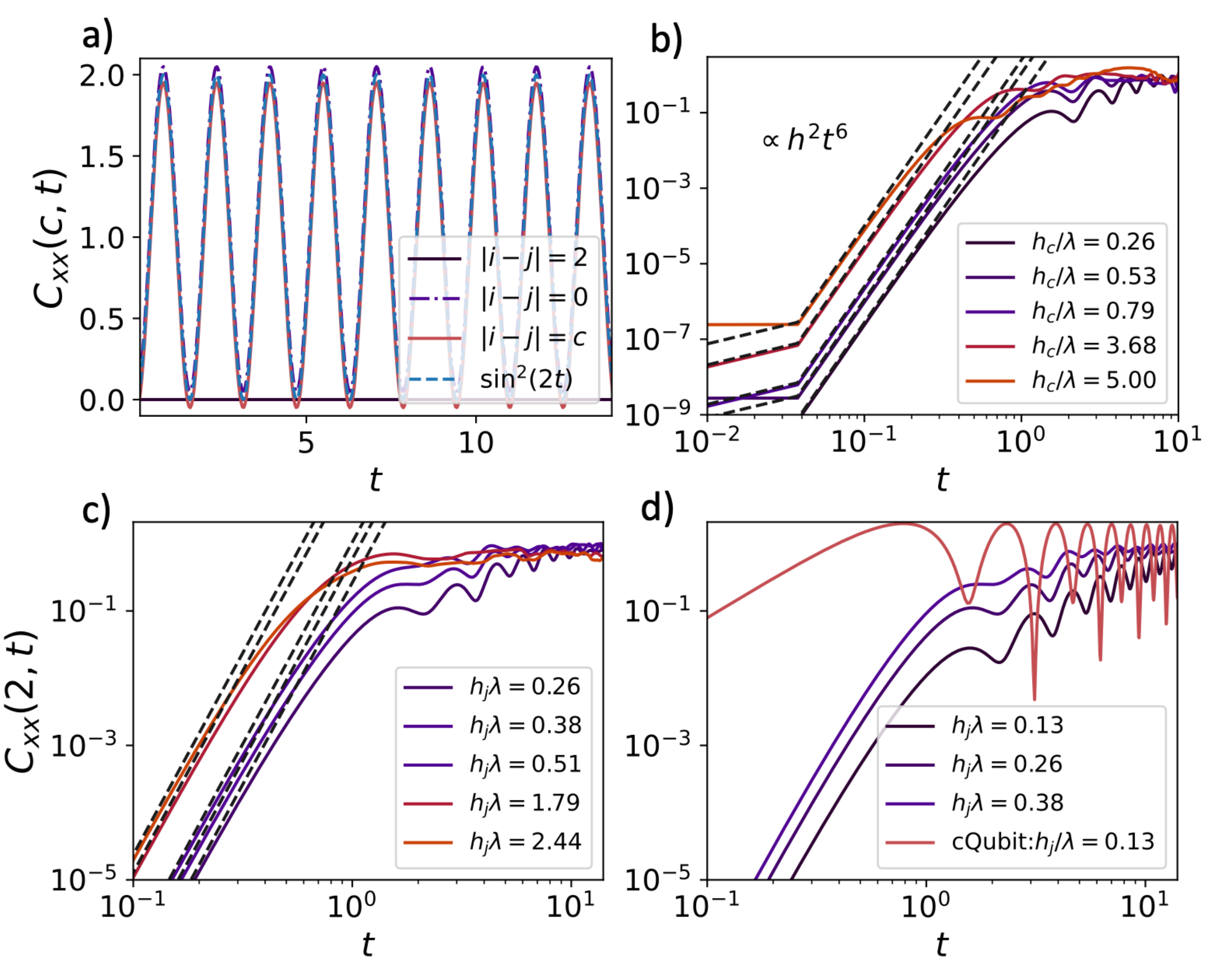}
    \caption{OTOC $C_{xx}(i,j,t)$ for (a) $h = 0$, (b) $h_c > 0$, (c, d) $h > 0$ with $L+1 = 13$, $\lambda = 1$. (a) OTOC is only nonzero on the edge and central c-qubit as coherent evolution under the star-coupling does not allow operators to propagate beyond the c-qubit. (b,c) early-time growth behavior is accurately captured by BCH expansion of the two-time correlator. (d) At intermediate times, $t$, as operator weight transitions back to the initial site, there is then essentially a waiting time that occurs before more operator weight may grow on sites $j \neq [i, 0]$. This leads to a polynomial growth in time of the two-time correlators.}
    \label{fig:star_otoc}
\end{figure}

In addition to the exactly solvable star-Ising model presented in the main text, we present how the dynamics extend with a magnetic field applied to the c-qubit and all other sites.
\begin{equation}
    H = \lambda \sum_i \sigma^z_c \sigma^z_i + h \sigma^x_i + h \sigma^x_c.
\end{equation}
The early-time behavior using the BCH expansion of the commutator and for general ($h/\lambda$).
\begin{align*}
    [\sigma^x_j(t), \sigma^x_i] &= \sum_m^\infty \frac{(it)^m}{m!}[[H, \sigma^x_i]_m, \sigma^x_j] \\
    \\
    [A, B]_m &= [A, [A,B]_{m-1}] ; [A, B]_0 = B \\
    \\
    [H, \sigma^x_i]_1 &= it[\sigma^z_i, \sigma^x_i] = it \lambda \sigma^z_0\sigma^y_i \\
    \\
    [H, \sigma^x_i]_2 = \frac{(it^2)}{2}[H, \lambda \sigma^z_0\sigma^y_i] &= \frac{-t^2}{2}(-\lambda^2 \sigma^x_i + h\lambda \sigma^y_0\sigma^y_i). \\
    \\
    [H, \sigma^x_i]_3 = \frac{(it^3)}{6}[H, ...] &= \frac{-it^3}{6}(-\lambda^3 \sigma^z_0\sigma^y_i + h\lambda^2 (\sum_{j \neq i} \sigma^x_0\sigma^z_j\sigma^y_i + \sigma^x_0\sigma^x_i) - h^2 \lambda \sigma^z_0\sigma^y_i) 
\end{align*}
In the limit $h/\lambda = 0, \lambda = 1$ we have:
\begin{align*}
    [\sigma^x_j(t), \sigma^x_i] &= [\sum_m \frac{(it)^{2m}}{2m!}\sigma^x_i + \frac{(it)^{2m+1}}{(2m+1)!}\sigma^z_0\sigma^y_i, \sigma^x_j] \\
    \\
    &= [\cos(2t)\sigma^x_i - i\sin(2t)\sigma^z_0\sigma^y_i, \sigma^x_j].
\end{align*}
Then the two-time correlator behaves as:
\begin{equation}
    \langle \sigma^x_i(t) | \sigma^x_i \rangle = \cos{2t}.
\end{equation}
And the OTOC goes as:
\begin{equation}
    C_{xx}(i, 1, t) = 2\sin^2(2t),
\end{equation}
strictly for $i \in {c,1}$. Under the operation of the central Ising interaction, the operator weight is completely preserved to live in the projective space as we have seen before. When adding a non-commuting term that rotates states away from this space, how does this then modify growth on the remaining sites? Looking back, we see that 
\begin{align}
    \langle \sigma^x_j(t) | \sigma^x_i \rangle \sim \mathcal{O}(1) \propto \frac{t^3}{6}h^2 \\
    C_{xx}(j,i,t) \propto h^2t^6.
\end{align}
As exhibited in Fig.\ref{fig:star_otoc}, we see that for fixed $\lambda = 1$, that $h^2t^6$ provides an excellent approximation to the early-time growth behavior of OTOC on next nearest neighbors to the initial site. This rapid growth on the central site and corresponding neighbors at early-times in essence captures the extreme early-time entanglement growth that is observed to grow steadily with growing $h/\lambda$. The interesting phenomena occurs when $t \sim \frac{\pi}{2\lambda}$, where in Fig.\ref{fig:star_otoc}(a) the operator space $q$ is completely preserved, allowing $C(0, i, t)$ to decay to zero. On this same timescale we see in (b, c) and most clearly in (d) that this leads to an effective wait time that is proportional $\lambda$ and similarly corresponds to the amount of operator weight outside of the space $q$ on the c-qubit. Also, any weight operator weight transferred to edge spins then undergoes similar oscillations seen in (a) during this wait time. In the weakly decoherent regime $h << \lambda$, the wait time leads to essentially linear OTOC growth with oscillations that similarly grow in time.

\section{Entanglement Spreading on the Star Graph}

\begin{figure}
    \centering
    \includegraphics[width=0.65\textwidth]{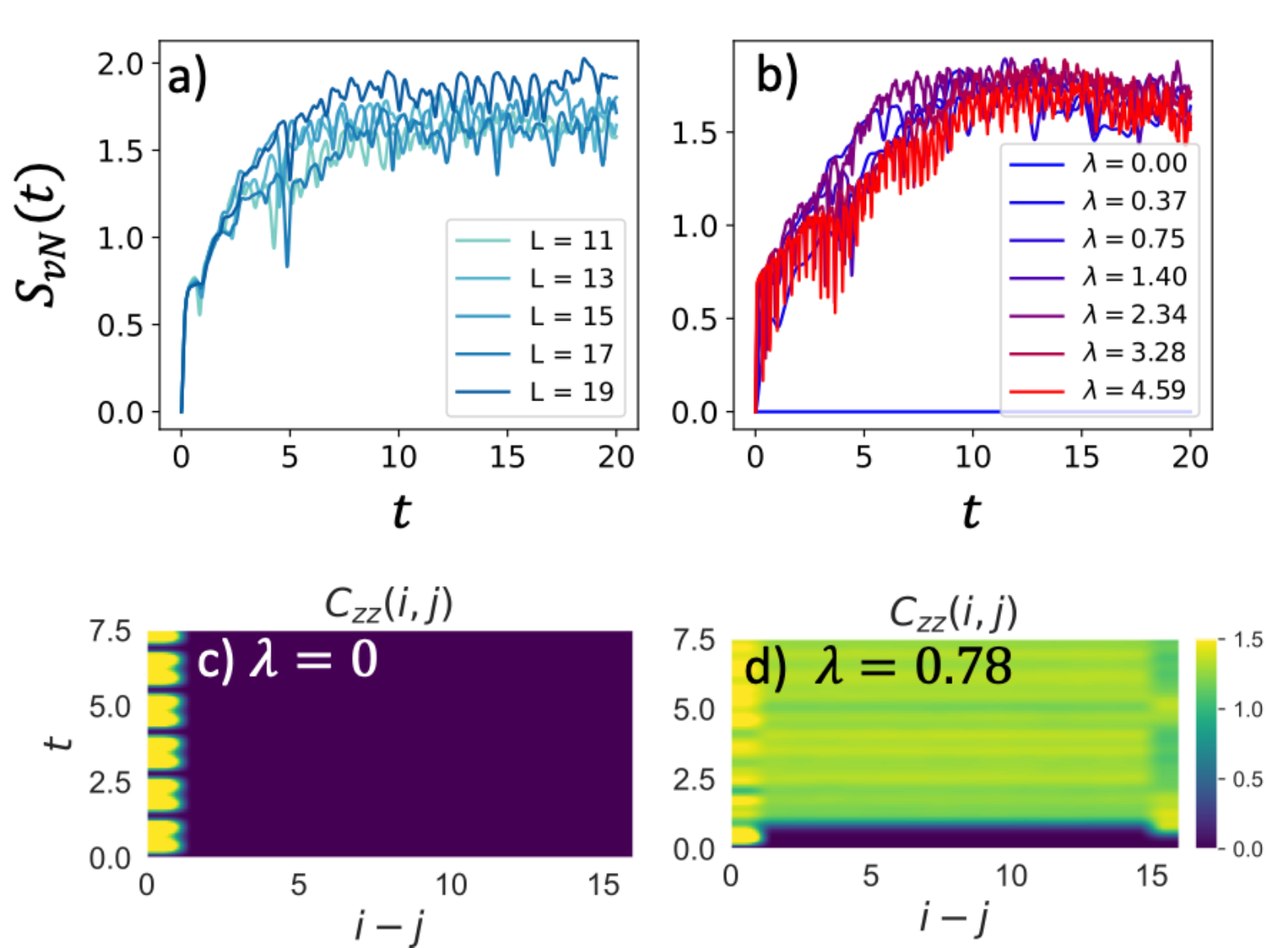}
    \caption{Information spreading star  model $(J = 0)$: Real-time evolution of the $1/2-$chain entanglement entropy initialized in $|+y\rangle$ (a, b), and OTOC following quench from a Haar random initial state. Information spreading in the star-Ising model $(J=0)$: Real-time evolution of the the $1/2-$chain entanglement entropy initialized in $|+y\rangle$ (a,b) and $C_{zz}(t, i-j)$ with initial state chosen randomly from the Haar distribution (c,d). (a) Entanglement entropy plotted as a function of length for $\lambda = 1.0$. Growth is independent of system size for fixed $\lambda$ and saturates to a value independent of system size. (b) Entanglement growth rate and steady state value for $\lambda > h$ saturate and become independent of $\lambda$. (c, d) OTOC $C_{zz}$ for $\lambda/\sqrt{L} = \{0, 0.78\}$, respectively. (c) Initial operator simply fluctuates under mixed-fields and does not propagate. (d) OTOC grows on the c-qubit at a rate that grows with $\lambda$ after an initial wait time that scales with $\text{min}[1/h, 1 / \lambda]$. Operator weight then spreads to the remaining leaves after a time $2 \ \text{min}[1/h, 1 / \lambda]$ corresponding to two processes of operator transfer that occur on the c-qubit.}
    \label{fig:star_rt}
\end{figure}

Here we examine the corresponding entanglement entropy growth on the Ising star graph. Starting from the same polarized product state used in the main results ($|+y\rangle$) we plot the half-system entanglement entropy following a quantum quench. We treat $\lambda$ as the variable and set $h, h_c = 1.05$ and $g, g_c = 0.45$. The entanglement entropy saturates to a length independent value Fig.\ref{fig:star_rt}(a), constant for system sizes $L = 11-19$. Though the mixture of onsite fields allows local operators to explore the operator space of single sites, the collective magnetic field and collective c-qubit interaction still limit the system to behave semiclassically. In Fig.\ref{fig:star_rt}(b) we note that the entanglement entropy growth rate and saturation value become independent for $\lambda > 1$. This is an interesting finding as we know in the large $\lambda >> h$ limit that operators simply scramble between the initial site and the cqubit. The corresponding entanglement entropy between halves is dominated by the system-qubit entanglement as entanglement generated between spins grows slowly. Once the system rapidly entangles with the cqubit, the entanglement entropy of the system remains fixed but becomes uniformly distributed between all sites. In the OTOC picture operators are scrambled with the cqubit and then leak into the remaining sites.

\section{OTOC Growth vs. System Size Scaling}

\begin{figure}
    \centering
    \includegraphics[width=0.55\textwidth]{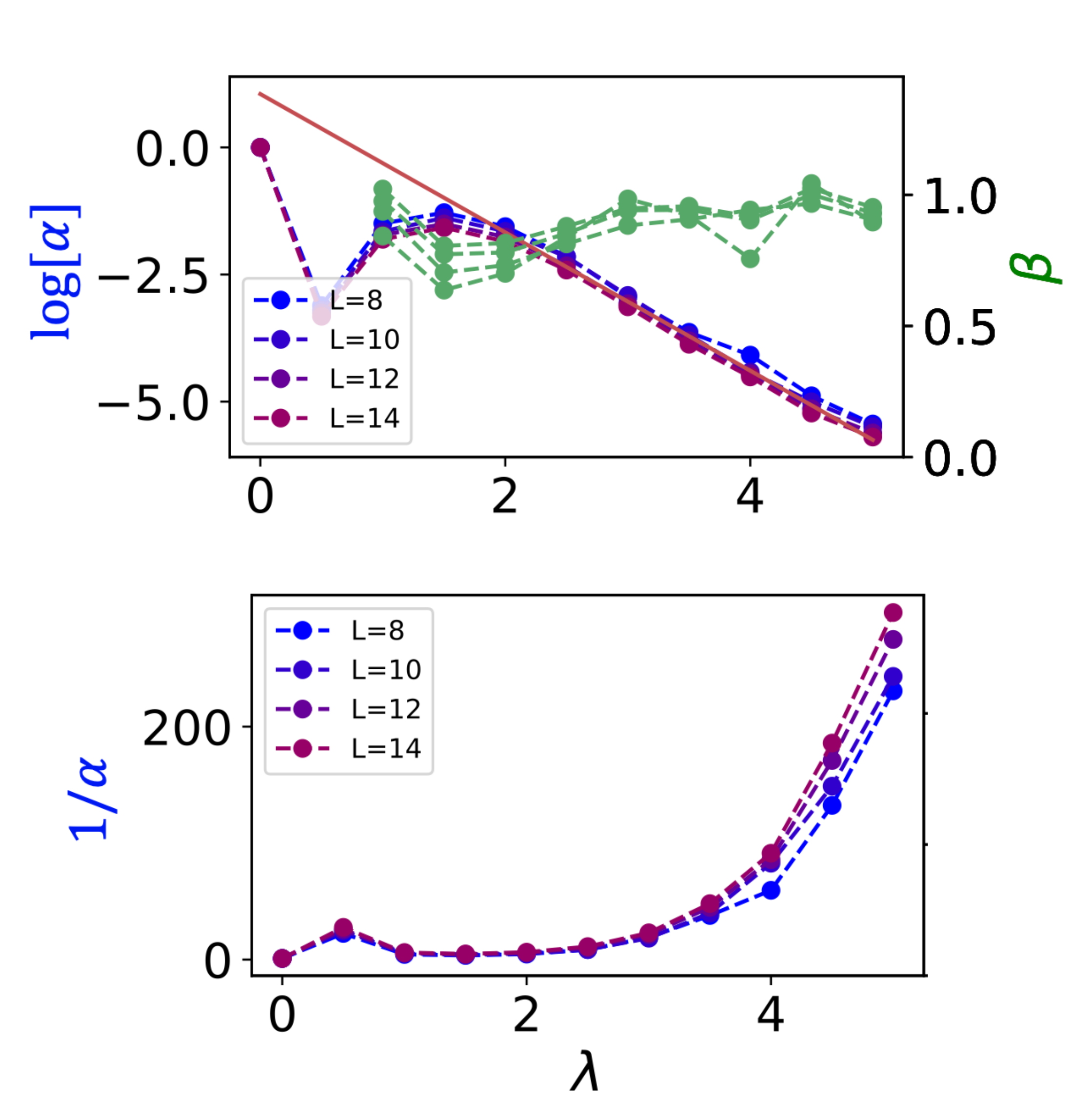}
    \caption{Finite size scaling of the coefficient of linear growth $\alpha t^\beta$ (see Main text Fig.\ref{fig:slow_growth}) in $C{zz}(|i-j| = 6, t)$ for system sizes $L = [8, 10, 12, 14]$. (Top) $\log[\alpha]$ decreases linearly with $\lambda$ as exponentially less operator weight decoheres on all sites and spreads throughout the system. (Bottom) Plot of $1/\alpha$ for better resolution of the system size dependence. We observe a small decrease in $\alpha$ with increasing system size, corresponding to the weak coefficient in $\lambda_c \sim \sqrt{L}$.}
    \label{fig:fs_slopes}
\end{figure}

In the main text we examined the longitudinal OTOC and found that it goes like $C_{zz}(t) \approx e^{-\lambda} t$. In Fig.\ref{fig:fs_slopes} we perform the same analysis and vary the system size. We see that close to the transition $\lambda \leq 1$ that the coefficient on OTOC growth is roughly independent of system size. For increasing $\lambda$ we see in Fig.\ref{fig:fs_slopes}(top) that there is weak decrease in $\log(\alpha)$ with system size. Looking at the $1/\alpha$ to highlight the slight difference in system size, we see that into the localized operator phase $\lambda > 3-4$ that larger system sizes exhibit increasingly slow operator growth.

\section{Spectral Statistics of the Star-Ising Model}

\begin{figure}
    \centering
    \includegraphics[width=0.725\textwidth]{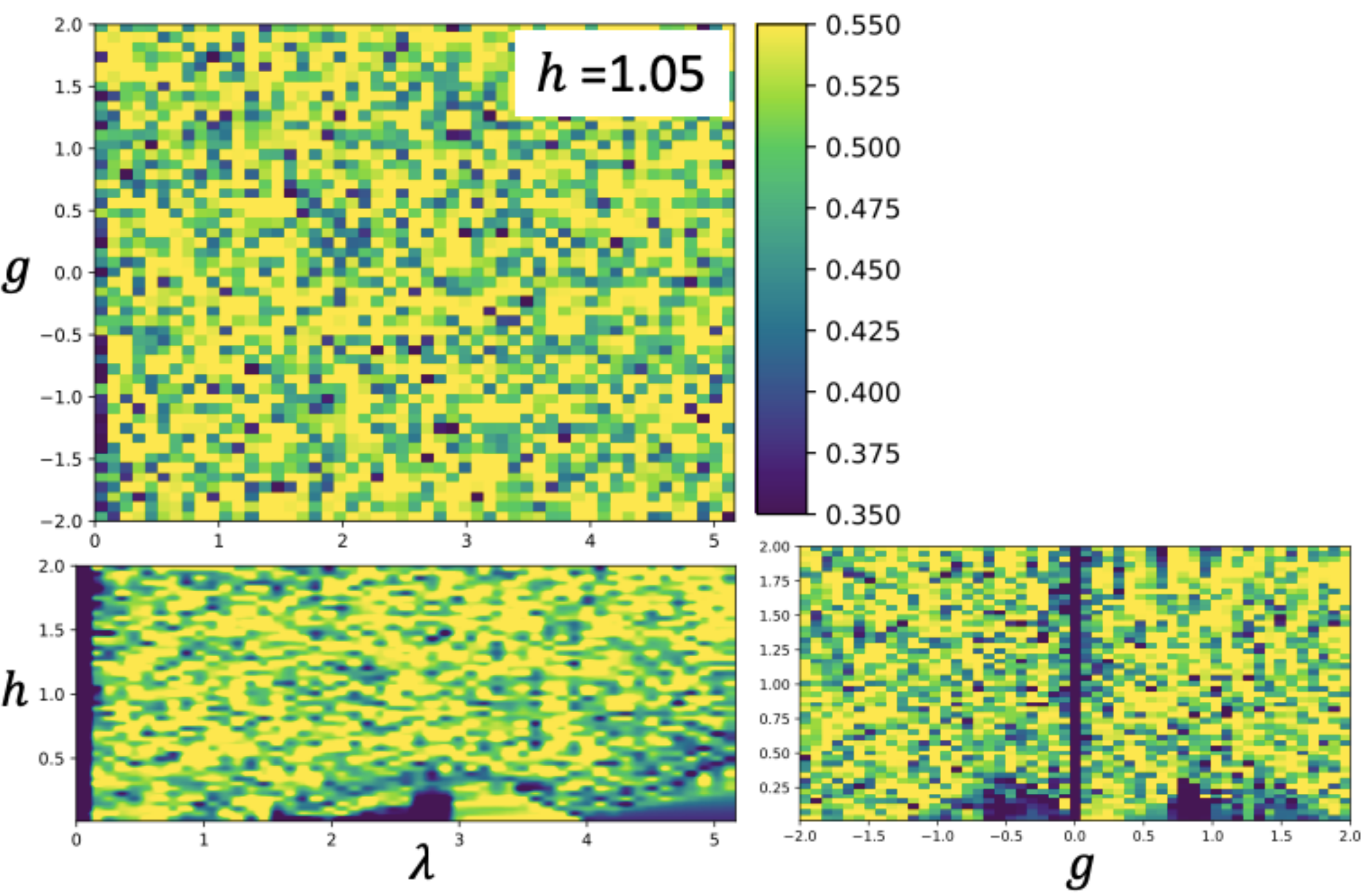}
    \caption{
    Average adjacency level spacing of energy eigenvalues $r(E_i, E_{i+1})$ for the ring-star Ising Hamiltonian for $L+1=16$ with periodic boundary conditions for fixed momentum and Z-reflection symmetry blocks of the Hamiltonian. (a) Spacing as a function of $\lambda, g$ for fixed $h$, (b) h, lambda, g = 0, and (c) $h, g, \lambda=0$. We find strong evidence of the integrable case of $h \neq 0, g = 0, \lambda = 0$ in (b,c).
    }
    \label{fig:energy_stats}
\end{figure}

We calculate the spectral statistics for the central qubit Ising model studied in the main text. These statistics provide insight into the random nature like behavior of the density matrix or whether there are conservation laws present. Anything beyond random statistics manifests as prethermalization, MBL, or integrable behavior when a system is quench from a highly energetic state. We calculate the spectral statistics using 100 states surrounding the middle of the energy spectrum. The ratio is between two levels $r_i$ is calculated as
\begin{equation}
    r_i = \text{min}[\frac{E_i - E_{i-1}}{E_i + E_{i-1}}, \frac{E_{i+1} - E_{i}}{E_i + E_{i+1}}]
\end{equation}
We then take the average spacing between all levels. We work in the $k = 0$ sector as well as the paritiy conserving sector, as the translational invariance leads to degeneracies between $k-$sectors when only one particular sector is meaningful under dynamics. We take the tensor product of this Ising spin-chain Hilbert space with the central qubit. In Fig.\ref{fig:energy_stats} we observe that the only point in phase space that does not exhibit GOE random matrix statistics $\langle r\rangle \approx 0.53$ as for $\lambda = 0, g = 0$, where we have the integrable TFIM. Here the system exhibits sub-Poissonian statistics.

\end{document}